\documentclass[twocolumn]{openjournal}
\usepackage{graphicx,amsmath,amssymb,amstext}
\usepackage{amsbsy,amsfonts,amsthm,color}
\usepackage[colorlinks,linkcolor=blue,citecolor=blue,urlcolor=blue ]{hyperref}
\usepackage[utf8]{inputenc}
\usepackage{float}

\defcitealias{Prole2023}{LP23}
\defcitealias{Schauer2021}{AS21}

\usepackage{graphicx}	
\usepackage{amsmath}	
\usepackage{amssymb}	
\usepackage{multirow}

\usepackage[utf8]{inputenc}
\usepackage[export]{adjustbox}
\usepackage{wrapfig}
\usepackage{dutchcal}
\usepackage{braket}
\usepackage{grffile} 
\usepackage{xspace} 
\usepackage{txfonts}

\usepackage{booktabs}

\usepackage{rotating}

\usepackage[framemethod=tikz]{mdframed}
\usepackage{lipsum}

\graphicspath{ {images/} }

\newcommand{\lya}{\mbox{${\rm Ly}\alpha$}\xspace}
\newcommand{\Lya}{Ly$\alpha$\xspace}

\newcommand{\lyart}{\texttt{LyaRT}\xspace}
\newcommand{\zelda}{\texttt{zELDA}\xspace}
\newcommand{\lasd}{\texttt{LASD}\xspace}

\newcommand{\igmz}{\texttt{IGM+z}\xspace}
\newcommand{\igm}{\texttt{IGM-z}\xspace}
\newcommand{\noigm}{\texttt{NoIGM}\xspace}
\newcommand{\recz}{\texttt{REC+z}\xspace}

\newcommand{\kms}{\,\ifmmode{\mathrm{km}\,\mathrm{s}^{-1}}\else km\,s${}^{-1}$\fi\xspace}
\newcommand{\vexp}{$V_{\rm exp}$\xspace}
\newcommand{\nh}{$N_{\rm H}$\xspace}
\newcommand{\ta}{$\tau_a$\xspace}
\newcommand{\ew}{$EW_{\rm in}$\xspace}
\newcommand{\w}{$W_{\rm in}$\xspace}

\newcommand{\mfa}{$\langle f_{\rm esc}^{4\AA} \rangle$\xspace}
\newcommand{\fa}{$f_{\rm esc}^{4\AA}$\xspace}
\newcommand{\mT}{$\langle T_{\rm IGM} \rangle$\xspace}

\newcommand{\zp}{\texttt{Z25}\xspace}

\newcommand{\wg}{$W_{\rm g}$\xspace}
\newcommand{\dl}{$\Delta \lambda_{\rm Pix}$\xspace}
\newcommand{\sn}{$S/N_p$\xspace}

\begin{document}


\title{Disentangling the galactic and intergalactic components in 313 observed Lyman-alpha line profiles between redshift 0 and 5}
    
   \author{Siddhartha Gurung-López$^{*,1,2}$}
   \author{Chris Byrohl $^{3}$}
   \author{Max Gronke $^{4,5}$}
   \author{Daniele Spinoso $^{6}$}
   \author{Alberto Torralba$^{1,2}$}
   \author{Alberto Fernández-Soto $^{8}$}
   \author{Pablo Arnalte-Mur $^{1,2,7}$}
   \author{Vicent J. Martínez $^{1,2,7}$}
   \email{$^*$email: gurung.lopez@gmail.com}
   \affiliation{1 - Observatori Astron\`omic de la Universitat de Val\`encia, Ed. Instituts d’Investigaci\'o, Parc Cient\'ific. C/ Catedr\'atico Jos\'e Beltr\'an, n2, 46980 Paterna, Valencia, Spain}
   \affiliation{2 - Departament d’Astronomia i Astrof\'isica, Universitat de Val\`encia, 46100 Burjassot, Spain}
    \affiliation{3 - Universität Heidelberg, Institut für Theoretische Astrophysik, ZAH, Albert-Ueberle-Str. 2, 69120 Heidelberg, Germany} 
    \affiliation{4 - Astronomisches Rechen-Institut, Zentrum für Astronomie, Universität Heidelberg, Mönchhofstraße 12-14, 69120 Heidelberg, Germany}
    \affiliation{5 - Max Planck Institute for Astrophysics, Karl-Schwarzschild-Str. 1, 85748 Garching, Germany }
    \affiliation{6 - Department of Astronomy, Physics Building, Tsinghua University, 100084 Beijing, China }
    \affiliation{7 - Unidad Asociada ``Grupo de Astrof\'{\i}sica Extragal\'actica y Cosmolog\'{\i}a'', IFCA-CSIC/Universitat de Val\`encia, Val\`encia, Spain}
    \affiliation{8 - Instituto de F\'{\i}sica de Cantabria (CSIC-UC), Avda. Los Castros s/n, 39005 Santander, Spain}


\begin{abstract}
Lyman-Alpha (\lya) photons emitted in star-forming galaxies undergo complex radiative transfer through the interstellar (ISM), circumgalactic (CGM), and intergalactic medium (IGM), imprinting characteristic signatures on their observed line profiles. We use the open-source package \texttt{zELDA} (redshift Estimator for Line profiles of Distant Lyman-Alpha emitters) to disentangle the galactic and intergalactic contributions in 313 \lya spectra observed with HST/COS and MUSE, spanning $0<z<6$. \texttt{zELDA} employs artificial neural networks trained on mock \lya spectra generated with Monte Carlo radiative transfer through thin-shell models and IGM transmission curves from the TNG100 simulation. We find that sources at $z<0.5$ exhibit minimal IGM attenuation, whereas at $z>3$ the IGM significantly suppresses the blue peak of \lya. After correcting for IGM effects, the stacked intrinsic galactic \lya line profiles display remarkably little evolution from $z=0$ to $z=6$. We measure the mean IGM \lya escape fraction, finding $\langle f^{4\AA}_{\mathrm{esc}}\rangle > 90\%$ for $z<0.5$, decreasing from $\sim0.85$ at $z=3$ to $\sim0.55$ at $z=5$. Our measurement of the redshift evolution of the \lya IGM escape fraction agrees with independent constraints on the IGM mean optical depth.  After a comparison between our  $\langle f^{4\AA}_{\mathrm{esc}}\rangle$ estimation and the global \lya escape fraction from the literature, our findings indicate that the IGM might dominate \lya observability at redshift z$\gtrsim$5.0, after which ISM and CGM effects tend to dominate at lower $z$. Our results demonstrate that \texttt{zELDA} enables robust reconstruction of intrinsic \lya spectra and provides a direct probe of the interplay between galactic outflows and IGM transmission across cosmic time.
\end{abstract}


\maketitle

\section{Introduction}\label{sec:intro}

\begin{figure*} 
        \includegraphics[width=7.2in]{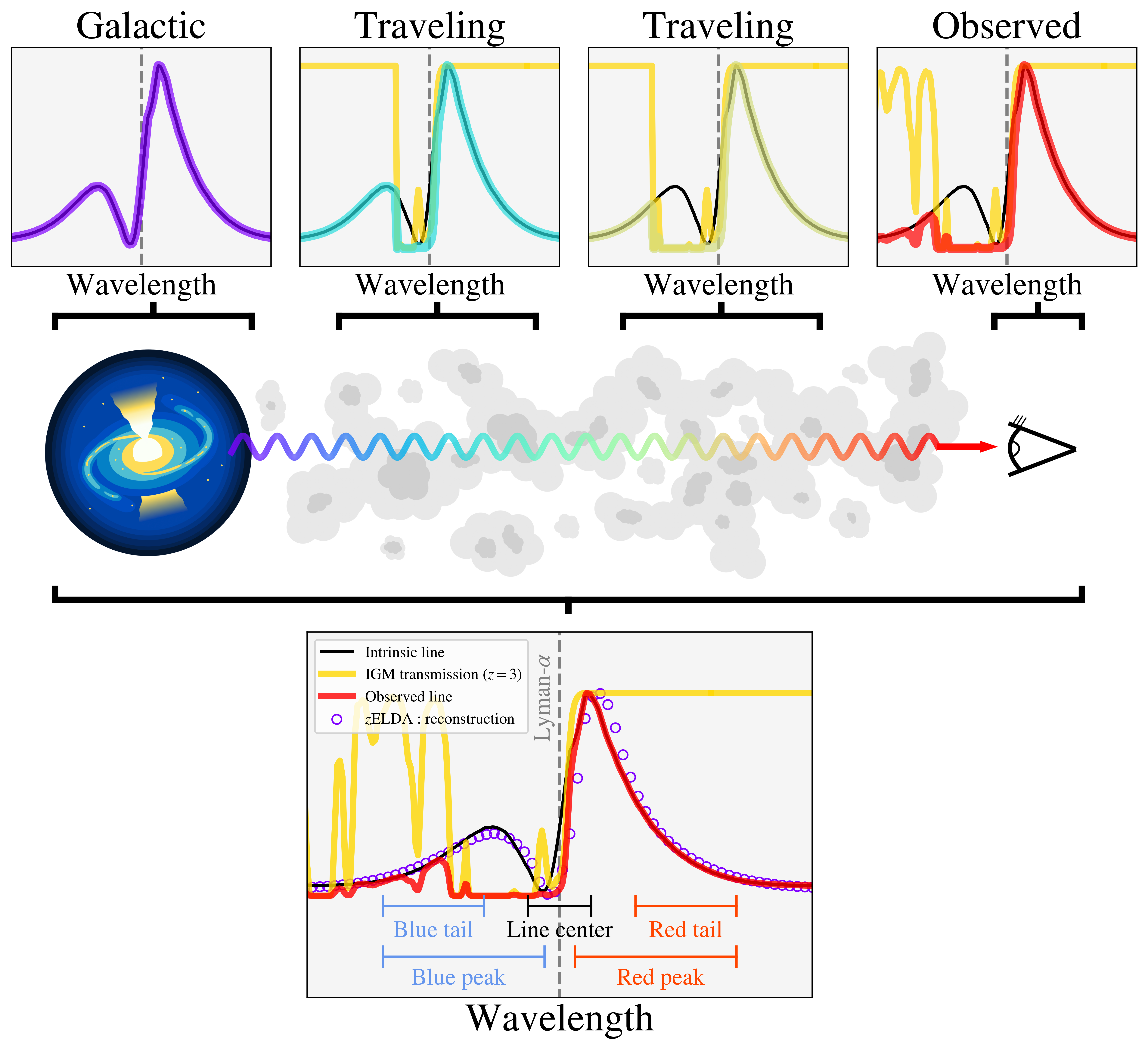}%
        \caption{Illustration of the impact of the IGM attenuation in the galactic \lya line profile. Through all the  top panels, we show the intrinsic \lya line profile emerging from the galaxy in thin black and the IGM transmission curve in yellow. In the top panels we show the galactic line profile obscured intrinsically (top left panel, purple), the same profile after traveling a short distance in the IGM (blue, second top panel) as well as further through the IGM (green, third top panel), and its shape when it  reaches the observer (red,top right panel). The bottom panel shows the \zelda reconstruction (purple) using as input the observed line profile (red), along with the line structure terminology used in this work. }
        \label{fig:mean_illustration}
        \end{figure*}

The first spectral line of neutral hydrogen, Lyman-$\alpha$ (\lya), plays a pivotal role in extragalactic astronomy \citep{OuchiObservationsLymana2020}. It is not only used to detect and characterize the most distant galaxies \citep[e.g.][]{2015ApJ...810L..12Z,witstok24} but thanks to its resonant nature,  it probes the dim outskirts of galaxies \citep{steidel10,Wisotzki2016}.

The emergent \lya line profile results from complex radiative-transfer processes in the interstellar, circumgalactic, and intergalactic media (ISM, CGM, and IGM; see \citealt{Dijkstra_2019} for a review).  
Because of the large scattering cross-section of neutral hydrogen, \lya photons are absorbed and reemitted many times before escaping the galaxy. Each scattering event changes both the direction and the frequency of the photon according to the velocity distribution of the gas. Consequently, the observed \lya profile is strongly shaped by the geometry, density, and bulk motions of the gas.

The fact that \lya observables probe gas from the ISM to the IGM holds tremendous potential. As such, \lya spectra are frequently used to investigate the porosity of the ISM and as a proxy for ionizing photon escape \citep[e.g.,][]{Izotov_2016,2016ApJ...828...71D,2017AA...597A..13V}. Furthermore, the existence of a dominant red peak in the \lya spectra \citep[e.g.][]{Erb_2014,Trainor_2015} is interpreted as a widespread presence of galactic outflows. 

On the scale of the CGM, \lya reveals the gas properties through the shape of the surface brightness profiles on scales of tens of kpc for individual \citep[e.g.,][]{Wisotzki2016} and even hundreds of kpc for stacked observations \citep{steidel11,2018Natur.562..229W,2022ApJ...929...90L}. In addition, thanks to the advent of integral field spectrographs such as \textit{MUSE} \citep{MUSE} or \textit{KCWI} \citep{KCWI}, one can now also study the emergent \lya spectrum changes as a function of projected position \citep[e.g.][]{Erb_2018,Leclercq_2017}, revealing, for instance, the transition from galactic out- to intergalactic inflows \citep{Erb_2018,Li_2022,Guo_2024}.

At even larger scales, the \lya photons emerging from the galaxy are shaped by the IGM, which contains neutral hydrogen in (mostly self-shielding) lumps.
Thus, further radiative transfer processes influence the \lya line profile. However, because photons interacting with the IGM gas change direction and are scattered out of the line of sight -- and the probability of scattering (back) into the line of sight is negligible -- one can treat these large-scale effects as effective absorption.
Importantly, in this process, photons are redshifted progressively by the Hubble flow as they travel through the IGM. This redshifting causes the IGM to attenuate higher frequencies of the galactic spectrum as photons advance \citep[e.g.][]{Laursen2011}.

We illustrate this in the top panels of Fig.\ref{fig:mean_illustration}. The galactic \lya line profile is shown in black, and the IGM transmission is drawn in yellow. In the second top panel (blue), we show a line profile after traveling a short distance in the IGM, which is slightly absorbed. Then, in the upper third top panel (green), we show the exact line profile after continuing travel through the IGM, which shows more attenuation. Finally, in the top right panel (red), we show how the line reaches the observer.

The fact that the IGM also changes the \lya line shape and the observability of \lya emitting galaxies leads to the fact that \lya emission at $z\gtrsim 5$ is now regularly used as a powerful tool to constrain the `Epoch of Reionization' \citep[for a review, see,][]{dijkstra14}. While predominantly the (non-)visibility, as well as the observed strength of the \lya emission (through the change of the luminosity function, the equivalent width distribution, or through the clustering of LAEs), is used to constrain the (global) neutral fraction, it is also possible to use the observed spectral shape to constrain the neutral gas of the IGM. For instance, the most blueward observed flux can be used to constrain the size of the ionized region around a given galaxy \citep{Mason_2020,Tang_2024}, in particular the visibility of an observed blue peak \citep{Matthee_2018,Gronke_2020}.

While this ability of \lya to probe neutral hydrogen on the ISM, CGM, and IGM scales provides the community with a potential treasure trove of information, this highly multiscale process naturally comes with many degeneracies. For instance, does the absence of a blue peak imply an ISM outflow or a neutral IGM? 
To unravel this degeneracy, we implemented the ability to reconstruct the transmission and the `galactic' \lya spectra (shaped by the ISM and CGM) into the Python package \zelda \footnote{Github: \url{https://github.com/sidgl/zELDA_II}. Tutorials and documentation on \zelda\ are also available at \url{https://zelda-ii.readthedocs.io/index.html}. } (redshift Estimator for Line profiles of Distant Lyman-Alpha emitters , \cite{zelda_25a}).
To fit observed \lya line profiles, \zelda is now able to disentangle between the galactic and the IGM components of a \lya line profile through artificial neural network models trained with mock \lya spectra (see \cite{zelda_25a} for a detailed description and testing of the algorithm or Sect.~\ref{sec:methodoogy} for a summary).


This work aims to demonstrate on real data, i.e., actually observed spectra, how one can disentangle between the IGM and the ISM/CGM effects shaping the \lya spectral shapes. With this separation of effects at hand, we will also study the evolving impact of the IGM on spectral shapes, or vice versa study the IGM evolution purely using \lya spectra. A similar exercise has been done in previous work using stacked line profiles \citep{Hayes11}. The novelty of this work is that  this will be the first time doing so using individual spectra and estimating the \lya escape fraction in the separate sources. While the current work represents only the first step on utilizing this potential treasure trove of information (and thus, is to some degree a proof-of-concept study), other applications such as spatial tomography of the IGM are thinkable.

This work is structured as follows: in Sect.~\ref{sec:methodoogy}  we briefly introduce \zelda's pipeline for recovering the galactic component from IGM attenuated \lya line profiles.  We introduce the observational data set used in this work in  Sect.~\ref{ssec:results_observations_data}. Next, we apply \zelda's methodology to the observational data set in Sect.~\ref{sec:results_observations} and measure the mean IGM \lya escape fraction as a function of redshift. In Sect.~\ref{sec:val_and_disucssion} we validate our results by comparing them to the literature. Finally, we draw our conclusions in Sect.~\ref{sec:conclusions}. 

In this work, we draw the intergalactic transmission curves and \lya line profiles in the rest frame difference to the \lya wavelength ($\Delta \lambda _0$). Redshift accuracy is provided in the same units. Typically, this quantity is expressed in velocity units, instead, as $\Delta v = c \Delta\lambda_0/\lambda_{\rm Ly\alpha}\approx(247\kms)\times \Delta\lambda / 1$\AA{}, where $c$ is the speed of light in vacuum and $\lambda_{\rm Ly\alpha}\approx 1215.67$\AA{} is the restframe \lya wavelength. 

Furthermore,  we refer to the \lya IGM escape fraction of a source as the ratio between the flux emerging from the ISM/CGM and that observed after the IGM along the observer's line of sight. This definition can be understood as the transmission of IGM around \lya and through the line of sight,  convolved with the line profile shape. In general, in the IGM, \lya photons are not destroyed by dust grains, although they are scattered out of the line of sight. This effectively causes a reduction of the observed \lya flux of the galaxy, while producing a diffuse/faint glow of the IGM  \citep{Byrohl_2023}.   

\section{Reconstructing attenuated Lyman-$\alpha$ emission lines}\label{sec:methodoogy}

In this section, we describe our approach to reconstructing observed \lya line profiles attenuated by the intergalactic medium and to estimating their \lya IGM escape fraction. Our methodology is based on the Python package \zelda. Here, we briefly summarize \zelda's pipeline, which is fully detailed in \cite{zelda_25a}, hereafter \zp.

To fit the observed \lya line profiles, we use the open-source Python package \zelda \citep{GurungLopez_2019b,gurung_lopez_2021, gurung_lopez_2022}. \zelda’s thin-shell model is implemented via linear interpolation within a precomputed grid of line profiles generated by the full Monte Carlo radiative transfer code \lyart \citep{orsi12}. The grid contains 3,132,000 nodes spanning a five-dimensional parameter space: bulk expansion velocity \vexp $\in [0,1000]\,\mathrm{km\,s^{-1}}$, neutral hydrogen column density \nh $\in [10^{17},10^{21.5}]\,\mathrm{cm^{-2}}$, dust optical depth \ta $\in [0.0001,1.0]$, intrinsic equivalent width \ew $\in [0.1,1000]\,\AA$, and line width \w $\in [0.01,6]\,\AA$ of the \lya emission before entering the thin shell. Further details on the grid and its accuracy can be found in \cite{gurung_lopez_2022}. 

The IGM transmission curves used to train the \igmz and \igm models are described in detail in \cite{Byrohl2020,Byrohl_21}. These were computed within the TNG100 simulation \citep{Naiman_2018,Nelson_2019,Marinacci_2018,Pillepich_2018,Springel_2018}. The \lya radiative transfer was calculated using an updated version of the \texttt{ILTIS} code \citep{Behrens_2019,Byrohl2020,Byrohl_21}. For each halo more massive than $5\times10^9 M_{\odot}$, transmission curves were obtained along 1000 random sightlines, at six redshift snapshots ($z=0.0,\,1.0,\,2.0,\,3.0,\,4.0,\,5.0$). These lines of sight integrate the neutral hydrogen optical depth from the halo radius ($\sim 1\,$cMpc) out to tens of cMpc. This  integration path ensures that it covers a sufficient distance for the Hubble flow to shift the \Lya emission out of resonance, thereby capturing all the relevant attenuation produced by neutral hydrogen in the IGM.

\zelda incorporates three models based on artificial neural networks (ANNs). Each model takes as input an observed \lya line profile. Both \igmz and \igm account for IGM attenuation and aim to recover the unattenuated best-fitting shell model. \igmz is trained to capture the redshift evolution of the IGM according to TNG100, while \igm is constructed to be redshift-independent. These two models therefore complement each other when measuring redshift-dependent properties. The third model, \noigm, completely ignores IGM effects and fits a shell model directly to the observed line profile.

The three artificial neural network models are trained with mocked \lya line profiles produced with \zelda and the IGM transmission curves from \cite{Byrohl2020,Byrohl_21}. {{ The mock line profiles cover homogeneously the outflow parameters of the shell model.} The training sets cover a wide range of spectral quality, from  spectral resolution but measured in wavelength in the observed frame  of \wg=0.1\AA{} to \wg=4.0\AA{} and signal to noise of the peak of the line between \sn=5.0 and \sn=20.0. \igmz, \igm, and \noigm use a different training set and input. In summary:

\begin{itemize}
    \item \igmz: The training set includes mock line profiles attenuated by the IGM. The IGM transmission curve associated with a particular \lya line profile uses the actual redshift of the source. By construction, the IGM evolution of \cite{Byrohl2020,Byrohl_21} is directly imprinted on the training set. Thus, \lya line profiles at lower redshift are less attenuated than those at high redshift. Finally, \igmz includes a proxy redshift in the input.
    \item \igm: The training set also includes mock line profiles attenuated by the IGM. Although, in contrast to \igmz, we use the re-calibrated. As shown in \zp,  \cite{Byrohl2020} IGM transmission curves were fine-tuned so that they match the mean evolution of the redshift of the IGM transmission (\mT) by \cite{Faucher-Giguere08}. Thus, the IGM transmission curves used in \igmz and \igm are slightly different. 
    transmission curves. These are drawn randomly without considering the mock source redshift. Thus, there is no evolution of the IGM through redshift in this training set. There will be \lya basically unattenuated (typical of $z=0$) and greatly attenuated (typical of high $z$) at all redshifts. Finally, unlike \igmz, \igm does not include the proxy redshift in the input.
    \item \noigm: The training set does not use IGM transmission curves. \noigm includes a proxy redshift in the input, as in \igmz. 
\end{itemize}

All three models provide the same type of output: the shell-model parameters with uncertainties, the estimated redshift, and the \lya IGM escape fraction within $\pm2\,\AA$ of the line center, \fa (for \igmz and \igm). By definition, \fa is obtained as the ratio of the IGM-attenuated \lya flux to the intrinsic flux. Since the IGM at $z<4$ primarily affects the blue side of \lya, \fa specifically measures absorption in the $\pm2\,\AA$ window, and thus, while related, differs from the mean IGM transmission.

The use of both \igmz and \igm enables us to test the robustness of our measurements. \igmz encodes the expected redshift evolution of IGM transmission, whereas \igm does not. Comparing them allows us to identify potential biases in redshift-dependent trends, such as spurious variations in the IGM escape fraction induced by the training set. If \igmz predicts a redshift trend absent in \igm, the effect is likely artificial and it could be attributed to our training set rather than being authentic. For reference, we also include results from \noigm, which ignores IGM attenuation entirely. 

In short, each of the ANNs (\igm and \igmz) takes as input the observed Ly$\alpha$ line profile, which has been sequentially shaped by radiative transfer in the ISM, CGM, and IGM. This input spectrum is shifted to a proxy rest-frame based on the wavelength of its peak flux. The ANN then outputs the true systemic redshift, the IGM escape fraction, and the underlying parameters of the thin-shell model. By evaluating these predicted thin-shell parameters, we generate what we refer to as the "reconstructed spectrum". This reconstructed profile represents the intrinsic shape of the Ly$\alpha$ line after it has escaped the dense galactic environment (ISM and CGM), but before undergoing large-scale intergalactic attenuation.

This paper aims to evaluate how the \lya IGM escape fraction evolve with redshift. While the mean IGM transmission is well known to evolve, the redshift dependence of the \lya escape fraction has not been directly measured before. On the other hand, \mfa depends on the local IGM environment (typically denser and more neutral than the field) as well as on the \lya line profile shape. Thus, it is unclear whether the \igmz training set reproduces its evolution accurately. We therefore include the \igm model as a redshift-independent control. 

When the spectral quality is sufficiently high, both \igm and \igmz recover \fa and the outflow parameters reliably, as shown in \zp. At low quality, however, the models behave differently due to their training sets. In particular, as quality decreases, the ANNs tend to return values that reflect prior distributions rather than true inference. Since \igmz is forced to encode redshift evolution, it may reproduce trends even when absent in the data, while \igm will not. The close agreement between their predictions in the results section is reassuring, as it indicates that the information content of the observed spectra is sufficient for accurate inference rather than being dominated by training priors.

We note that our models do not explicitly assume a redshift evolution for the intrinsic properties of the galaxies (e.g., morphology). Because the mapping between physical galactic properties and effective thin-shell parameters is not tightly constrained, we instead use a highly broad training set. This unconstrained approach prevents artificial biases and allows the ANN to empirically recover any potential redshift evolution of the outflow parameters directly from the data.

It is important to emphasize that the thin shell model is a physical simplification. We use it primarily as an effective tool to span a wide diversity of Ly$\alpha$ line profile shapes rather than as a literal representation of galactic geometry. While linking the effective thin shell parameters to true macroscopic galaxy properties is complex and non-trivial, the reconstruction of the intrinsic line shape and the IGM escape fraction remains robust as long as the true profile can be  described by the model. Indeed, out of $\sim 350$ available LASD spectra, only $\sim 11$ sources (as described in Sect.~\ref{ssec:results_observations_data}) could not be reproduced by our implementation and were excluded. Reproducing these specific sources would likely require models explicitly including clumpiness or asymmetrical escape channels derived from high-resolution hydrodynamical simulations, which we leave for future work.

 The ANN takes as input the observed Ly$\alpha$ line profile, which has been sequentially shaped by radiative transfer in the ISM, CGM, and IGM. This input is shifted to a proxy rest-frame using the line's peak flux. The ANN then outputs the true systemic redshift, the IGM escape fraction  and the parameters of the thin-shell model. Throught the thin shell parameters, which are recover without the IGM attenuation, we build the intrinsic o reconstructed line profile. When we refer to the "reconstructed spectrum", we denote the pre-IGM line profile generated using the shell parameters predicted by the ANN. This represents the intrinsic shape of the Ly$\alpha$ line after escaping the dense galactic environment but before large-scale intergalactic attenuation. 

\section{ Description of observational data   }\label{ssec:results_observations_data}

The observational \lya line profiles used in this work are stored in the {\it Lyman Alpha Spectral Database} \citep[\texttt{LASD},][]{Runnholm_2020}\footnote{\url{http://lasd.lyman-alpha.com}}, which currently contains 348 \lya emission lines in the redshift range $0<z<6.0$. We exclude from our analysis a total of 35 ($\sim 10\%$) sources, all these are shown in Fig.~\ref{fig:LASD_bad}. We removed sources in which:
\begin{enumerate}
    \item \lya line profiles with a very steep continuum around \lya ,  since in our shell model we assumed a flat continuum,  (e.g. sources 0 and 23 in Fig.~\ref{fig:LASD_bad}),
    \item spectra with low \sn,   (e.g. sources 6, 20, 22 and 29  in Fig.~\ref{fig:LASD_bad}),
    \item spectra in which the blue peak dominated the red peak, as we only considered outflows in our shell model (e.g. sources 4, 5, 19 and 25  in Fig.~\ref{fig:LASD_bad}). 
    \item \zelda is trained with single source \lya line profiles, thus we excluded spectra that apparently contain \lya emission from multiple sources (e.g. sources 3 and 6  in Fig.~\ref{fig:LASD_bad}). 
\end{enumerate}
After these cuts, our final sample comprises 313 \lya line profiles spanning $0<z<6.0$. All spectra are shown in Appendix~\ref{app:lines}.
        
    The \lya line profiles from \texttt{LASD} were obtained mainly from two experiments: { i)} the Cosmic Origins Spectrograph \citep[{\it COS}][]{Green_2012} on board the {\it Hubble Space Telescope} ({\it HST}) and { ii)} the MUSE-WIDE survey \citep{Urrutia2019A&A...624A.141U,Herenz2017}. \lya line profiles obtained via {\it HST} were observed in the General Observers (GO): GO 11522 and 12027 \citep[PI: Green,][]{salzer_2001,Wofford_2013}, GO11727 and 13017 \citep[PI: Heckman,][]{Heckman_2011,Heckman_2015}, GO 12269 \citep[PI: Scarlata,][]{Songaila_2018}, GO 12583  \citep[PI: Hayes][]{Hayes_2014,Rivera-Thorsen_2015}, GO12928  \citep[PI: Henry][]{Henry_2015}, GO 13293 and 14080 \citep[PI: Jaskot][]{Jaskot_2014,Jaskot_2017}, GO 14201 \citep[PI: Malhotra][]{Yang_2017} and GO 13744 \citep[PI: Thuan][]{Izotov_2016,Izotov_2018,Izotov_2020}.

    HST-COS and MUSE-WIDE exhibit very different spectral quality. The spectra obtained by HST have excellent spectral quality ($R\sim 16,000$). In particular, the \lya line profiles exhibit excellent wavelength sampling, as \dl might take two particular values: 0.0598\AA{} and 0.0735\AA{}, depending on the medium-resolution gratings G130M and G160M. The spectral resolution is also of high quality. \wg  ranges from 0.073\AA{} to 0.10\AA{} with median 0.085\AA{}. The signal-to-noise ratio of the maximum of the line, \sn, spans a wide range from \sn=6.5 to $\sim 400$ with a median of $\sim$38 and only a $\sim11\%$ of the sample exhibit \sn $<15.0$. Considering MUSE data, the spectral quality is more limited ($R\sim 3000$), \dl=1.25\AA{}. Meanwhile, \wg changes from 1.25\AA{} at $z=3$ to 4.0\AA{} at $z=6.0$. We only consider \lya line profiles such as \sn$>$5.0. The maximum \sn is 35.0, and the median is 7.8. 

\begin{figure*} 
        \includegraphics[width=7.2in]{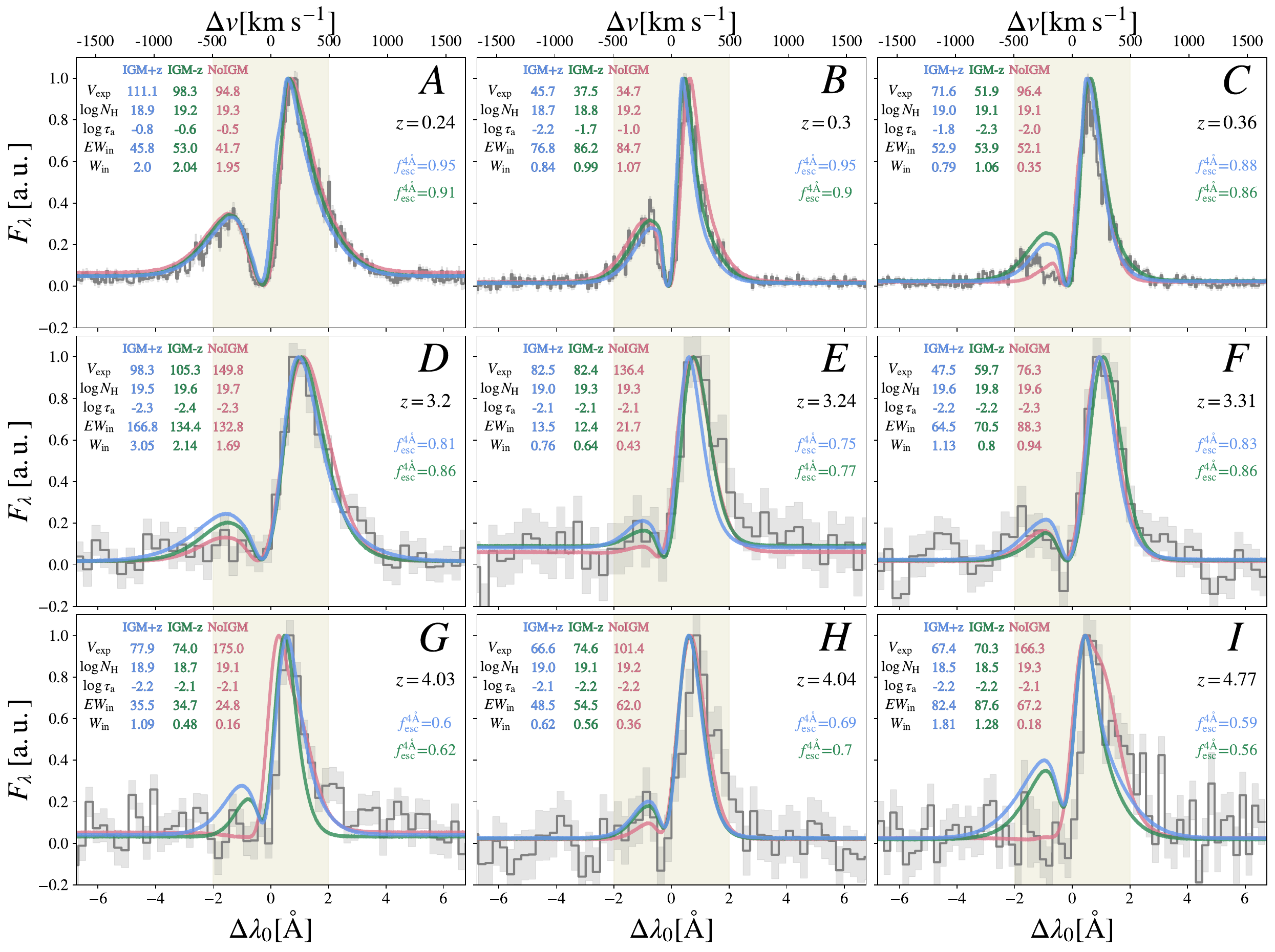}
        \caption{Nine examples of \zelda's prediction on line profiles displayed in the rest frame. The observed line profile is shown in dark grey with its $1\sigma$ uncertainty in light grey. \zelda's reconstruction using the models \igmz, \igm, and \noigm is displayed in blue, green, and pink, respectively. The redshift of the source is shown in the top right corner. \zelda's estimations of \fa given by the \igmz and \igm models are shown in blue and green, respectively. The central grey-shaded region shows the wavelength interval of \fa. The estimated outflow parameters are indicated in the table with each subplot. \vexp is given in \kms, \nh in $cm^{-2}$ and \ew and \w in \AA{}. All the fitted line profiles are also shown in Appendix \ref{app:lines}.}
        \label{fig:LASD_examples}
        \end{figure*}

    In \zp, different spectral-quality configurations were explored. Most HST line profiles correspond to the best-case scenarios described there, while the MUSE spectra match intermediate and worst-case configurations. It was demonstrated in \zp that \zelda can successfully reconstruct stacked \lya line profiles for both MUSE and HST qualities, and can reliably track the redshift evolution of \fa. 

    The \sn thresholds applied here were chosen to ensure accurate recovery of the stacked line profile, \fa, and its mean \mfa across redshift, as shown in \zp. Specifically, for HST sources \zelda recovers the stacked line profile with a Kolmogorov–Smirnov (KS) estimator better than 0.04.\footnote{The KS estimator is computed between the reconstructed and true intrinsic \lya line profiles for 10,000 mock spectra spanning the full range of \zelda outflow parameters, IGM absorption levels, and spectral-quality configurations.} For MUSE sources the stacked line profile is recovered with KS $<0.11$. As shown in \zp (Sect.~4.3), with the spectral qualities of both MUSE and HST the redshift evolution (or lack thereof) of the stacked \lya profile is recovered robustly. Regarding \mfa, for HST data both \igmz and \igm retrieve unbiased values within 3\%, whereas for MUSE-quality spectra \mfa may be biased by up to 10\%. The precision of \fa estimates is evaluated using mock \lya profiles generated with \zelda under varying spectral qualities and IGM absorption levels.

\begin{figure*} 
        \begin{center}
        \includegraphics[width=4.8in]{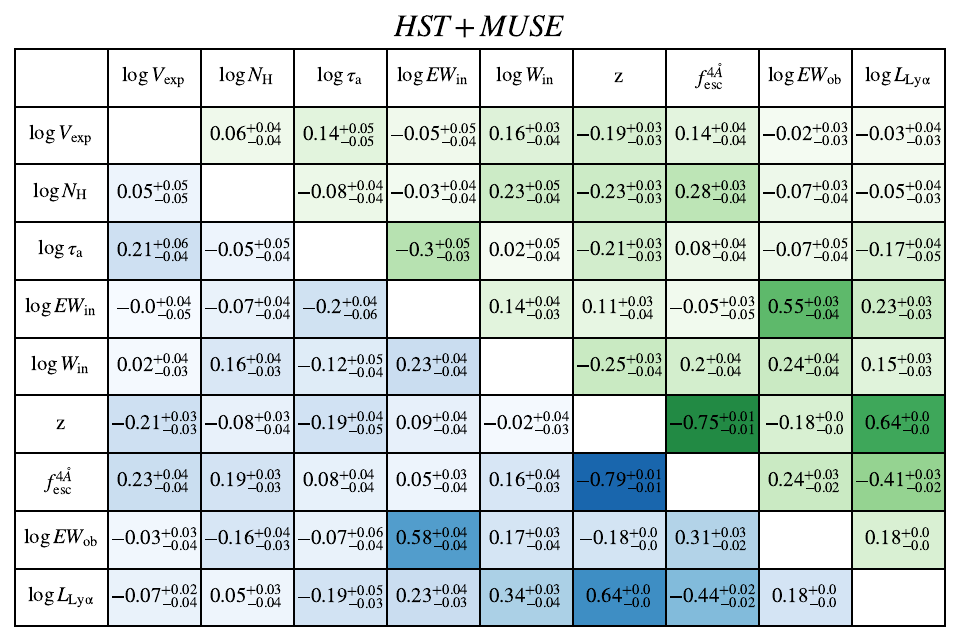}
        \end{center}
        \caption{  Spearman correlation coefficient and its 1$\sigma$ associated uncertainty for HST and MUSE sources ,for \igmz (bottom blue) and \igm (top green).  }
        \label{fig:spear_0_7} 
        \end{figure*}

\section{Results on observed Lyman-$\alpha$ line profiles}\label{sec:results_observations}


In this section, we apply \zelda's ANN models to observed \lya line profiles. In Sect.~\ref{ssec:results_observations_lines} we present a few examples of the predicted \lya line profiles.  We study the correlation between \fa and outflow properties in Sect.~\ref{ssec:results_observations_properties}. We present our observed \mfa reconstruction in Sect.~\ref{ssec:results_observations_fesc}.


\subsection{ Individual reconstructed \lya line profiles  }\label{ssec:results_observations_lines}

In Fig.~\ref{fig:LASD_examples}, we present nine examples illustrating the performance of \zelda on observed \lya line profiles. The observed spectra are shown in grey, while the predictions of \igmz, \igm, and \noigm are displayed in blue, green, and pink, respectively. The top row corresponds to HST sources ({\it A}, {\it B}, and {\it C}), whereas the middle and bottom rows show MUSE LAEs ({\it D}–{\it I}). 

Overall, the three ANN models reproduce the red peak of the observed \lya line profiles well. 
The Kolmogorov–Smirnov (KS) estimator, defined as the maximum difference between two cumulative distributions,  in the red peaks (defined as the wavelength interval from $\lambda_{\rm Ly\alpha}$ to $\lambda_{\rm Ly\alpha}$+5\AA in rest frame) has a median of $\sim 0.07$ and scatter of 0.11 (for more details see Sect.~\ref{ssec:line_profile_fitting_acc}).     Both \igmz and \igm also predict the presence of a blue peak, while \noigm tends to follow the observed spectra more closely without introducing additional features. This behavior is consistent across both HST and MUSE data.  Validation tests on mock spectra in \zp demonstrate that \zelda does not artificially generate blue peaks when intrinsically absent. Consequently, the blue peaks predicted by \igmz and \igm are data-driven features rather than artifacts of the shell-model priors.

At low redshift (top row of Fig.~\ref{fig:LASD_examples}), the three models predict nearly identical intrinsic(before IGM absorption but after ISM/CGM interaction) profiles for sources {\it A} and {\it B}, with close agreement between the predicted and observed spectra in both the red and blue peaks. This consistency reflects the negligible IGM attenuation at these redshifts. For source {\it C} ($z=0.36$), all models produce similar red peak shapes and amplitudes, but differ slightly in the blue peak: \igm predicts the strongest blue component, followed by \igmz and then \noigm. These differences correlate with the reconstructed expansion velocity, which decreases from 96.4\kms\ (\noigm) to 71.6\kms\ (\igmz) and 51.6\kms\ (\igm). This trend suggests that including IGM attenuation impacts the inferred kinematic parameters, with \igm providing a more accurate fit by incorporating IGM constraints.

At higher redshift (middle and bottom rows), IGM absorption becomes more significant. For sources {\it D}, {\it E}, {\it G}, {\it H}, and {\it I}, both \igmz and \igm predict a reconstructed blue peak—attenuated or partially absorbed—whereas \noigm either predicts no blue peak or one with much lower amplitude. For instance, in source {\it I}, the suppressed blue peak predicted by \igmz and \igm (\fa$\sim0.56$) matches the observations, while \noigm fails to reproduce this feature. This highlights the importance of IGM modeling for accurately reconstructing line profiles at high redshift. Moreover, the small difference in \fa values (within 5\%) between \igmz and \igm further underscores the robustness of their predictions.

The observed and reconstructed \lya line profiles of all the sources are shown in Appendix \ref{app:lines}.  Fig.~\ref{fig:LASD_0} for sources with $0.0<z<0.20$, Fig.~\ref{fig:LASD_1} for sources with $0.20<z<0.35$, Fig.~\ref{fig:LASD_2} for sources with $0.35<z<3.3$ , Fig.~\ref{fig:LASD_3} for sources with $3.3<z<3.7$ , Fig.~\ref{fig:LASD_4} for sources with $3.7<z<4.2$ ,Fig.~\ref{fig:LASD_5} for sources with $4.2<z<4.8$ , Fig.~\ref{fig:LASD_6} for sources with $4.8<z<6.0$. In general, the red peak of the \lya predicted by the ANN models follows the observations closely. The blue peak predictions follow the observations at low redshift, while at high redshift, they diverge for \igmz and \igm.

\begin{figure*} 
        \includegraphics[width=7.2in]{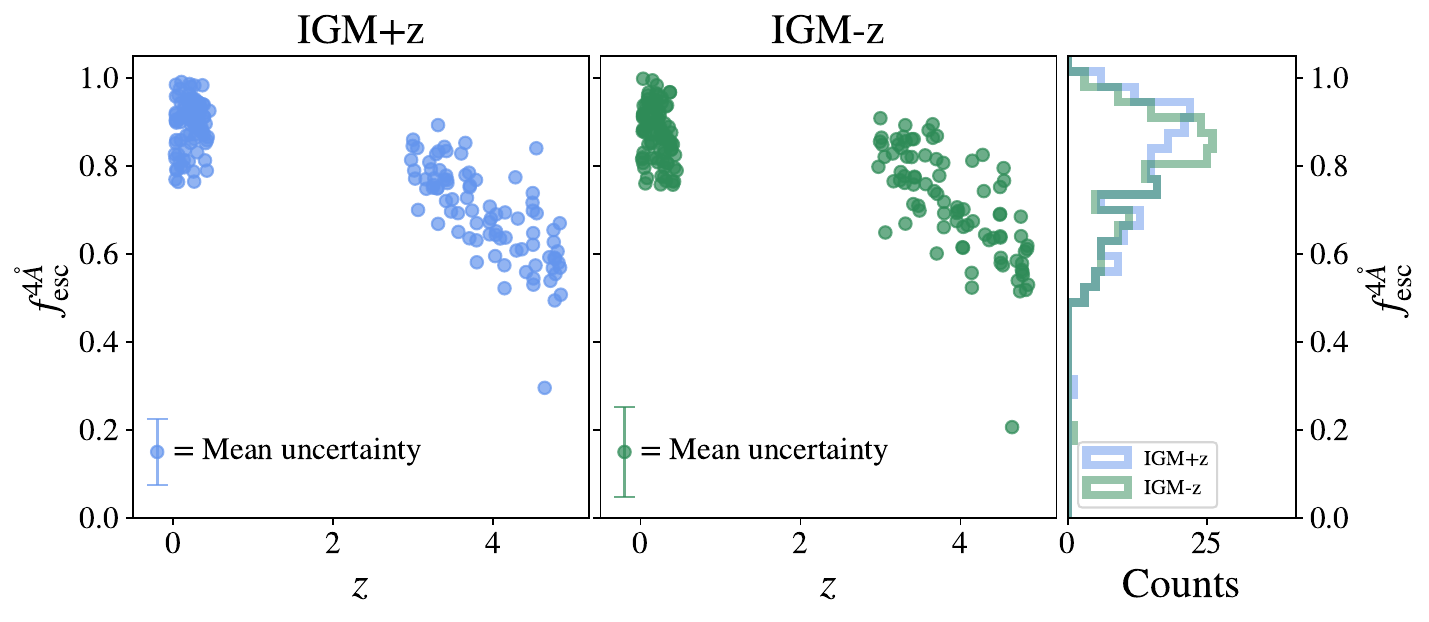}%
        \caption{ \fa as a function of redshift for \igmz (left) and \igm (middle). The right panel shows the histogram of the predicted \fa for \igmz and \igm in blue an green, respectively. }
        \label{fig:f_esc_z_comp}
        \end{figure*}

\subsection{ Correlations between model and observed properties  }\label{ssec:results_observations_properties}

In this section, we investigate correlations between the thin-shell outflow properties predicted by \zelda, redshift, the Ly$\alpha$ luminosity $L_{\rm Ly\alpha}$, and the rest-frame observed Ly$\alpha$ equivalent width $EW_{\rm ob}$. The latter two quantities are taken directly from \lasd.

We quantify the strength of the possible correlations by the Spearman correlation coefficients for the full sample, which are shown in Fig.~\ref{fig:spear_0_7}, along with their $1\sigma$ uncertainties. Meanwhile, the full 1-dimensional and 2-dimensional distributions can be found in Appendix~\ref{ssec:prop_1d_2d_distributions}. The coefficients from \igm are shown in the top-right corner (green, $\rho$), and those from \igmz in the bottom-left corner (blue, $\rho_z$). Uncertainties are estimated following \cite{gurung_lopez_2022}: we draw 1000 realizations of the outflow parameters from the ANN posteriors (see methodology in \zp), compute the Spearman correlation coefficient for each realization, and define the $1\sigma$ interval from the 16th and 84th percentiles of the resulting distribution. We consider two properties to be correlated (or anti-correlated) if the absolute value of the Spearman coefficient exceeds 0.45. Correlation matrices for the HST and MUSE subsamples separately are provided in Appendix~\ref{app:corr_muse_hst}.  

Overall, we find that the correlation coefficients inferred from \igmz and \igm are very similar: both models typically identify the same correlations.  In Appendix~\ref{app:corr_muse_hst}, we present the coefficients for the low-redshift (HST, left) and high-redshift (MUSE, right) samples (see Fig.~\ref{fig:spear_1_7}). Across the full dataset, only a handful of parameters show significant correlations, some of which are present in both subsamples (low and high redshift), while others appear only in one.  These relations are  the following: 

\begin{itemize}
    \item \textit{Correlation between \ew and $EW_{\rm ob}$:}  
    The intrinsic equivalent width (\ew) correlates strongly with the observed equivalent width ($EW_{\rm ob}$), particularly at low redshift. The correlation weakens toward higher $z$, likely due to increasing IGM absorption of the Ly$\alpha$ line. Within the MUSE sample, the correlation strength declines further with redshift, reflecting the growing impact of IGM absorption.

    \item \textit{Redshift and Luminosity Relation:}  
    A clear relation is found between redshift and Ly$\alpha$ luminosity: HST sources at low redshift exhibit lower $L_{\rm Ly\alpha}$ than MUSE sources at higher redshift. This is likely a selection effect, as fainter sources are preferentially detected at low $z$.

    \item \textit{Anti-correlation between \fa and Redshift:}  
    We find a strong anti-correlation between the Ly$\alpha$ IGM escape fraction (\fa) and redshift. \fa remains nearly constant for $z \lesssim 0.5$, but declines sharply from $z \sim 3$ to $z \sim 6$. This trend is consistent with independent measurements from Ly$\alpha$ forest studies, confirming the increasing role of the IGM in attenuating Ly$\alpha$ emission at high redshift \citep[e.g.][]{Faucher-Giguere08}.

    \item \textit{Correlation between \nh and \fa:}  
    A moderate correlation is observed between the neutral hydrogen column density (\nh) and \fa in the MUSE sample, which is weaker for HST sources. This suggests that high-redshift sources with higher \nh may have broader Ly$\alpha$ line profiles, potentially reducing scattering events in the IGM and leading to higher escape fractions.

\end{itemize}

\begin{figure*} 
        \includegraphics[width=7.1in]{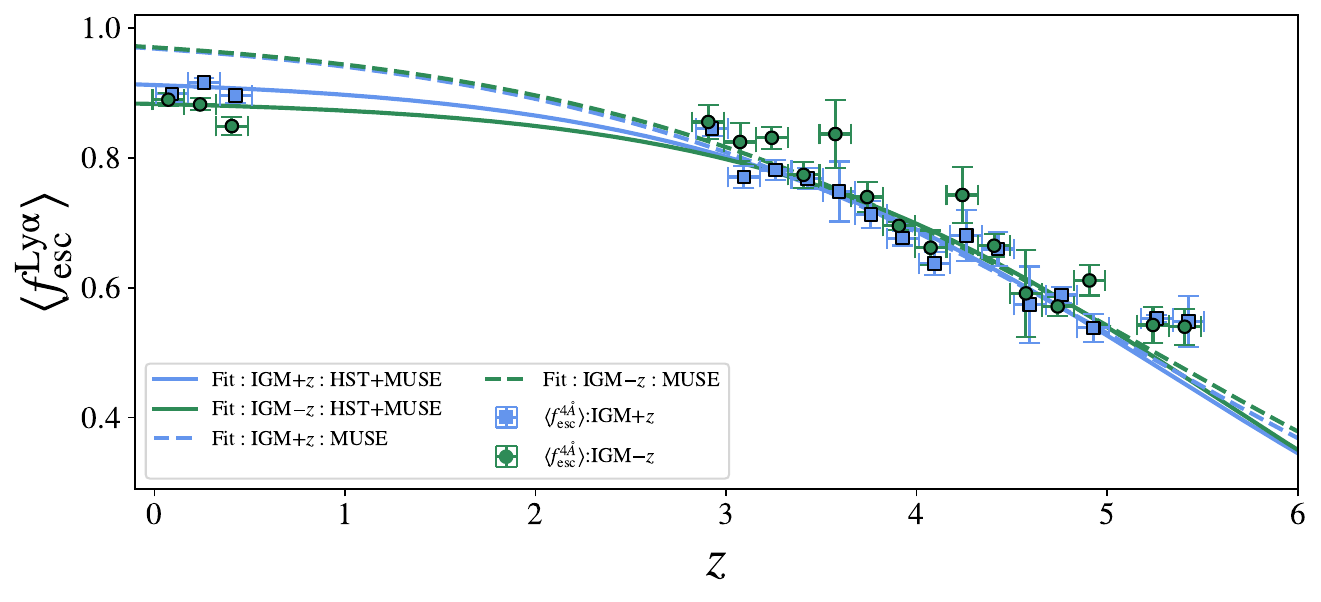}%
        \caption{ Mean \fa as a function of redshift for \igmz (blue squares) and \igm (green dots). The solid lines indicate the fits, including all the sources below $z=5$. Meanwhile, the colored dashed line indicates the fit using only MUSE data below $z=5$.  }
        \label{fig:f_esc_z_mean}
        \end{figure*}


\begin{table*}
\caption{Best-fitting parameters for the mean \lya IGM escape fraction (\mfa) as a function of redshift, modeled with the modulated Fermi-Dirac distribution given by Equation~\ref{eq:1}.}
\begin{center}
\begin{tabular}{lccccc}
\hline
$z$ range & Experiments & ANN model & $a$ & $b$ & $c$ \\ \hline
 0.0-5.0  &  HST+MUSE  &  IGM+$z$  &  5.35$\pm$0.0143 &  0.8$\pm$0.0084  &  0.93$\pm$0.0002 \\
 0.0-5.0  &  HST+MUSE  &  IGM-$z$  &  5.5$\pm$0.0366 &  0.86$\pm$0.0196  &  0.89$\pm$0.0002 \\
 2.9-5.0  &  MUSE      &  IGM+$z$  &  5.18$\pm$0.187 &  0.66$\pm$0.084  &  1.0$\pm$0.033 \\
 2.9-5.0  &  MUSE      &  IGM-$z$  &  5.25$\pm$0.326 &  0.66$\pm$0.133  &  1.0$\pm$0.057 \\ \hline
\end{tabular}
\end{center}
\label{tab:fitting_parameters}
\end{table*}

\subsection{IGM escape fraction evolution with redshift  }\label{ssec:results_observations_fesc}

In this section, we examine in detail the redshift dependence of the IGM escape fraction, \fa, as predicted by the models that include an IGM component (\igmz and \igm).  

As shown in \zp (Sect.~4.3 and Appendix C), the accuracy of \fa depends strongly on the spectral quality of the \lya line profiles. We therefore restrict our analysis to spectra with \sn$>8.0$ and \wg$<3.0 $\AA{}. This selection yields 93 HST and 113 MUSE sources, for a total of 206 \lya line profiles.  These cuts ensure that the uncertainty in \fa remains below 0.05 for HST sources and below 0.09 for MUSE sources.

Fig.~\ref{fig:f_esc_z_comp} shows the \fa values for these 206 sources as obtained with \igmz (left) and \igm (middle), as well as their overall distributions (right). In both ANN models, \fa decreases with increasing redshift. For HST sources ($z<0.55$), \fa is concentrated between 0.8 and 1.0, whereas for MUSE sources at $z>3.0$, values range between 0.85 and 0.5, with a typical dispersion of $\sim0.2$. This trend is consistent with expectations: the higher neutral hydrogen content and increased IGM density at larger redshifts lead to stronger Ly$\alpha$ absorption, reducing the escape fraction.  

In Fig.~\ref{fig:f_esc_z_mean}, we present the mean  \lya   emission line IGM escape fraction, \mfa, for \igm (blue squares) and \igmz (green squares). Uncertainties correspond to the error of the mean, and the bins have been shifted by $\pm0.01$ in redshift for clarity. At $z<0.5$ (HST data), \igmz and \igm diverge slightly, with \igmz predicting a constant value of \mfa$ \sim0.91$ and \igm \mfa$ \sim0.89$.  At $z>3.0$ (MUSE data), however, both models agree closely, predicting that \mfa declines from $\sim0.9$ at $z=2.9$ to $\sim0.6$ at $z=5.0$. 

In Fig.~\ref{fig:f_esc_z_mean} we also show the best fitting of the predicted \mfa to a modulated Fermi-Dirac distribution (\zelda is able to recover the \fa-$z$ evolution as discussed in \zp), i.e., 

\begin{equation}\label{eq:1}
    \displaystyle \langle f_{\rm esc}^{4\AA} \rangle = \displaystyle\frac{c}{ e^{ b (z-a) } + 1 }.
\end{equation}

The best-fitting parameters are listed in Tab.~\ref{tab:fitting_parameters}. For the full redshift range (HST+MUSE), the fits yield $c=0.93$ for \igmz and $c=0.89$ for \igm. In the interval $2.9<z<5.0$, the parametric curves differ by less than 0.05. When restricted to MUSE data alone, both ANN models produce nearly identical fits, with $c=1.0$ and $a$ and $b$ consistent within $1\sigma$. This explains the differences at $z<2.0$ between the best-fitting curves derived from the full sample and from MUSE-only data. Extrapolating the fits, both models predict that the \lya IGM escape fraction approaches zero at $z\sim11$.

 As discussed in Sect.\ref{sec:methodoogy}, the simultaneous use of \igmz and \igm provides a powerful consistency check on our results. While \igmz encodes the expected redshift evolution of IGM transmission, \igm is designed to be redshift-independent. Any discrepancies between them would reveal potential biases in our training sets, for example spurious redshift trends in the IGM escape fraction. The fact that both models yield very similar predictions is therefore reassuring: it shows that the measurements are not driven by priors or training artifacts, but by genuine information contained in the observed spectra. This agreement strengthens the robustness of our conclusions regarding the redshift evolution of the Ly$\alpha$ IGM escape fraction. 


\section{ Validation and discussion}\label{sec:val_and_disucssion}

 In this section we aim to validate our results on the reconstructed \lya line profile  shapes and \mfa redshift evolution. This task is made complex by an inherent aspect of Lya line profile reconstruction: we can only observe one sight line per source. In \zp we calibrated and tested \zelda using multiple IGM transmission sight lines for the same \lya line profile, which is possible only for simulated spectra. We refer the reader to \zp for an in-depth analysis of the performance of \zelda on mock \lya spectra with HST and MUSE quality. In order to show the robustness of the \lya line profile reconstruction and \mfa estimation we performed the following tests:

\begin{itemize}
    \item We show that the inferred redshift is highly accurate ($\sim$0.3\AA{}) , with \igm being the least biased (see Sect.~\ref{ssec:results_observations_zaccuracy}). 
    \item We show that the red tail of the \lya line profile, supposedly unaffected by the IGM at low redshift,  is well recovered, specially  when implementing the IGM absorption (see Sect.~\ref{ssec:line_profile_fitting_acc}). 
    \item We computed the \lya stacked line profiles from the reconstructed \lya line profiles (see Sect.~\ref{ssec:results_observations_stack}). We find that the stacked \lya line profiles do not evolve strongly with redshift. This indicates that the evolution in the observed \lya spectrum arises from the IGM absorption, in agreement with previous findings in the literature \citep{Hayes2021,Hayes_2023}.
    \item We demonstrate that our \mfa measurement in individual \lya line profiles is compatible with the IGM mean optical depth found in the literature, especially at $z>2$ (see Sect.~\ref{sec:comparison_mfa_IGM}) .
    \item We find that our measurement of the IGM \lya escape fraction is consistent with the global \lya escape fraction reported in the literature in the whole redshift range (see Sect.~\ref{sec:disucssion_f_esc_global}).
\end{itemize}

Finally, we discuss the fact that the \igmz and \igm models predict that the \lya line profiles are IGM attenuated at $z\sim0$ in Sect.~\ref{sec:disucssion_igmz_at_low_z}.

\subsection{  Accuracy of redshift estimates with \zelda  }\label{ssec:results_observations_zaccuracy}

In this section, we evaluate the redshift accuracy of \zelda. For this purpose, we use HST sources with systemic redshifts measured from features other than \lya, yielding a total of 111 sources. We focus on HST data because MUSE sources in \lasd lack systemic redshift measurements below $z=0.5$. To quantify the precision, we analyze the distribution of rest-frame wavelength differences between the redshift predicted by \zelda ($z^{\rm zELDA}$) and the systemic redshift ($z^{\rm Sys}$), computed as in \cite{gurung_lopez_2022}:

\begin{equation}
        \Delta \lambda ^{\rm zELDA}_0 = \lambda _ {\rm Ly\alpha} \frac{z^{\rm zELDA}-z^{\rm Sys}}{ 1+z^{\rm Sys}},
    \end{equation}
as shown in Fig.~\ref{fig:z_accuracy}. The method estimates the systemic redshift with a rest-frame wavelength accuracy of $\sim 0.3\, \mathrm{\AA}$ (equivalent to a velocity uncertainty of $\sim 75\, \mathrm{km\,s^{-1}}$).

Our analysis shows that the three models \igmz and \igm and \noigm perform similarly. All of them recover accurately the \lya wavelength location in the line profiles with an uncertainty of $\sim 0.3$\AA{}. At low redshift, where the IGM has only a minor impact on the \lya line profile, the performance gain of \igmz and \igm over \noigm is modest. However, at higher redshifts the effect of the IGM becomes more significant, and we expect the IGM-inclusive models to substantially outperform \noigm. We note that the median uncertainty of the systemic redshifts in the HST sample corresponds to $\sim 0.39$\AA{} in the rest frame. Given that this value is comparable to the residuals we find, the measured dispersion is likely dominated by the uncertainties in the systemic redshift rather than the intrinsic precision of \zelda.

A small systematic bias is present across all models, with \zelda generally overestimating the \lya rest frame by approximately 0.25\AA{} (equivalent to $\sim$61\kms). This overestimation aligns with findings from previous studies, such as \cite{Orlitova_2018} and \cite{Li2021}, and points to inherent limitations of the `thin shell model.' It is noticeable that \igmz and \noigm are biased evenly while \igm shows a better performance. 

\subsection{ Line profile fitting accuracy  }\label{ssec:line_profile_fitting_acc}

\begin{figure} 
        \begin{center}
        \includegraphics[width=3.4in]{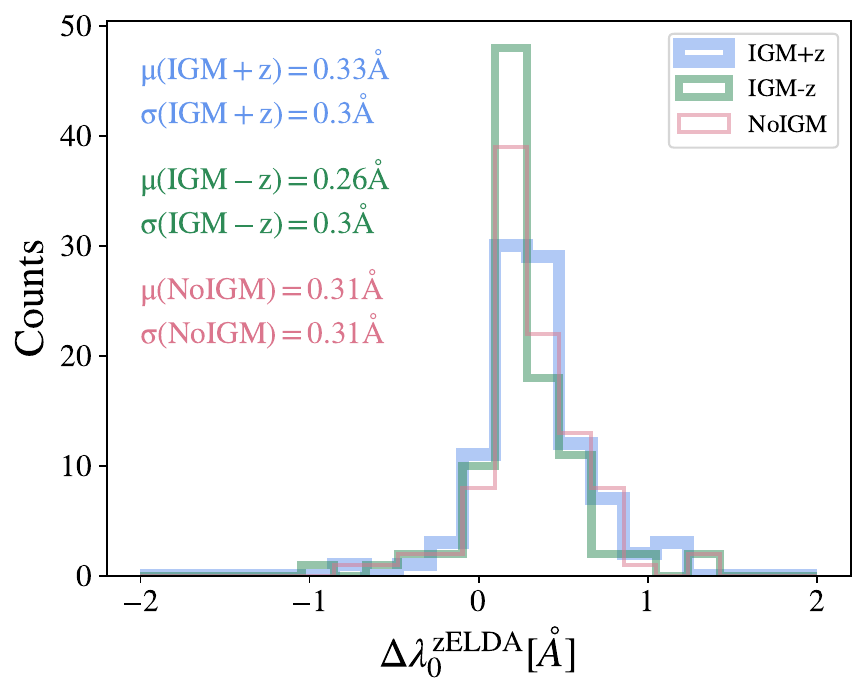}
        \caption{ Rest frame difference between \zelda's \lya wavelength and the estimated \lya wavelength using spectral features different from \lya in HST sources. \zelda's predictions are shown blue, green, and pink for \igmz, \igm, and \noigm, respectively. The mean and standard deviation of the distributions are shown for each model. }
        \label{fig:z_accuracy}
        \end{center}
        \end{figure}

\begin{figure} 
        \includegraphics[width=3.6in]{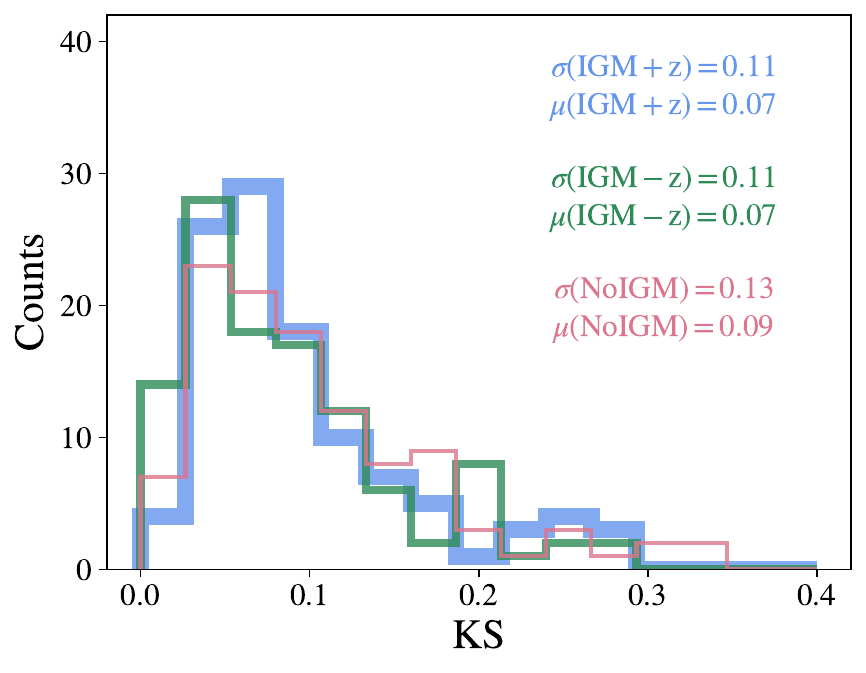}%
        \caption{ Left: Distribution of the Kolmogórov-Smirnov (KS) estimator in the red tail of HST \lya line profiles for \igmz, \igm, and \noigm in thick blue, medium green, and narrow pink, respectively. The width of the distribution is shown for each distribution, computed as the difference between the 84th and 16th percentiles and the median. }
        \label{fig:red_tails}
        \end{figure}

\begin{figure*} 
        \includegraphics[width=7.2in]{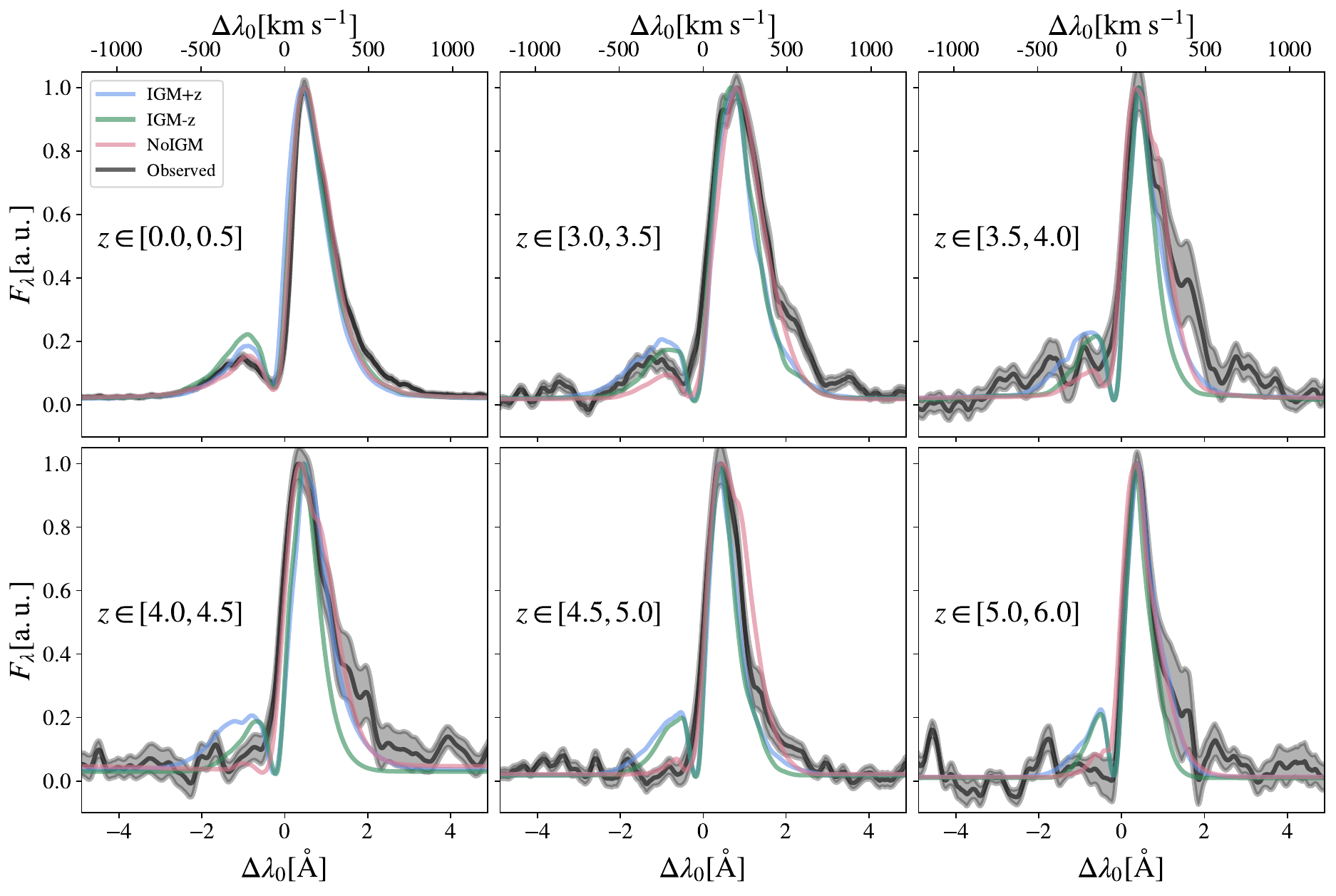}%
        \caption{ Comparison between stacked \lya line profiles. The observed stacked \lya line profile is shown in grey, while \zelda's prediction using \igmz, \igm, and \noigm are shown in blue, green, and pink. The stacked \lya line profile is computed in 6 redshift bins. The relevant redshift interval. The shaded dark regions marks the 1-$sigma$ uncertainty of the stacked line profile.}
        \label{fig:stack_comp}
        \end{figure*}

To compare the models fairly, we evaluate the accuracy with which they recover the intrinsic Ly$\alpha$ line shape in the observed spectra by focusing on the red tails. At low redshift, galaxies are expected to be only weakly affected by the IGM. Most of the absorption occurs blueward of Ly$\alpha$, leaving the red side largely unaffected. Consequently, the peak of the red component and its redward tail should closely resemble those of the intrinsic Ly$\alpha$ profile prior to IGM transmission. Comparing the red tail of the observed profiles with the predicted intrinsic spectra in the same wavelength range therefore provides a fair test of model performance. We define the red tail as the wavelength interval between the maximum of the red peak and $\lambda_{\rm Ly\alpha}$+ 5\,\AA in the rest frame of the source.

In Fig.~\ref{fig:red_tails}, we show the distribution of the KS estimator for the \igmz (blue), \igm (green), and \noigm (pink) in the red tails of the HST sources.  We find that \igm and \igmz outperform \noigm. In particular, the median of the KS distribution for the models is 0.07 for \igmz, 0.07 for \igm, and 0.09 for \noigm. Meanwhile, \noigm exhibits a little bit more dispersion than \igmz and \igm. We conclude that the \igmz and \igm models predict the red tail of the observed \lya line profile with the same precision.

\subsection{ Stacked Lyman-alpha line profile }\label{ssec:results_observations_stack}

In this section, we analyze the redshift dependence of the observed \lya stacked line profile shape (a similar analysis was done by \citealp{Hayes2021,Hayes_2023}; we focus here on the reconstruction of the pre-IGM spectra using \zelda).  We limit our analysis to \lya line profiles with \sn$>$9.0. This ensures that more than $\sim$85\% of the \lya line profiles of HST are recovered with KS$<$0.1. Meanwhile, as \w changes from 1.25\AA{} to 4\AA{} in MUSE data, $\sim$85\% and $\sim$65\% of the line profiles should be recovered with KS$<$0.1, respectively (see appendix D in \zp). \sn$>$9.0 was chosen as a compromise between reconstruction quality and number counts. We verify that similar results are obtained across different signal-to-noise thresholds (\sn$>$6.0, 7.0, 8.0, 10.0), with higher thresholds producing noisier stacked line profiles, while the overall trends remain consistent.

We compute the stacked \lya line profile from a sample of observed \lya line profiles as follows. First, we apply \zelda's ANN model (\igmz, \igm or \noigm) to the line profiles to obtain the reconstructed \lya line profiles and the redshift of the source. All the line profiles are shifted with respect to their rest frame in a fixed wavelength array  ranging from $\lambda_{\rm Ly\alpha}$-10\,\AA to $\lambda_{\rm Ly\alpha}$+10\,\AA in bins of 0.1\AA. The individual \lya line profiles are normalized so that their maximum reaches unity. Then, the stacked line profile is computed as the median flux in each wavelength bin. 

In Fig.~\ref{fig:stack_comp}, we present the \lya stacked line profiles across six redshift bins: $0.0<z<0.5$ (94 sources), $3.0<z<3.5$ (19 sources), and $3.5<z<4.0$ (17 sources) in the top row (left to right), and $4.0<z<4.5$ (10 sources), $4.5<z<5.0$ (13 sources), and $5.0<z<6.0$ (5 sources) in the bottom row (left to right). To ensure a fair comparison, the \lya profiles predicted by the ANN models are computed at the same resolution (\wg) as the observations.

In general, we find that the observed stack spectrum (shown in grey in Fig.~\ref{fig:stack_comp}) changes with redshift. At $0.0<z<0.5$ we find a faint but detectable continuum. Meanwhile, at higher redshifts, no continuum is detected. At $0.0<z<0.5$, the \lya stacked line profile exhibits a clear double peak line profile, with a prominent red peak that extends $\sim 2\AA{}$ redder than \lya. The blue peak is weaker and also extends $\sim 2\AA{}$ bluewards  \lya. The same structure of the double peak is also apparent at  $3.0<z<3.5$. 
Thus, the \lya stacked line profiles for $0.0<z<0.5$ and $3.0<z<3.5$ exhibit only minor differences.
 At $3.5<z<4.0$, the blue peak becomes less pronounced and eventually disappears for $z>4.0$, while the red peak remains consistently present across all redshift bins.
This is consistent with the findings by \citet{Hayes_2021}, {  i.e., these results are compatible with the scenario in which the \lya line profile is attenuated by the IGM progressively at higher redshifts and the pre-IGM spectra show little evolution. The IGM absorption at these redshifts mainly affects the blue side of \lya emission. This causes the main evolution in the \lya stacked line profile to be at the blue peak. }

Again, the observed stacked \lya line profile at redshift $0.0<z<0.5$ and that at $3.0<z<3.5$ exhibit a very similar blue peak. This is consistent with our measurements of the \mfa and the redshift invariance of the galactic stack line profile. We find that \mfa changes little from $z\sim0.5$ to $z\sim3.0$. Therefore, if the galactic stacked line profile is the same at $z\sim0.5$ to $z\sim3.0$, then the observed stack should be very similar at these redshifts. Reassuringly, this is what we find. {  This would imply that the IGM absorption in the \lya line profiles would not evolve strongly from $z\sim0$ to $z\sim3$. In contrast, the mean IGM transmission changes significantly in this redshift range. \mT evolves from $\sim1$ at $z=0$ to $\sim0.75$ at $z=3$ \citep{Faucher-Giguere08}. Thus, our results would indicate that the stacked \lya line profile is not very sensitive to the IGM evolution in this redshift window}.

Regarding the \lya stacked line profiles predicted by the ANN models, we find that all three (\igmz, \igm, and \noigm) generally reproduce the observed profile redwards of the \lya wavelength. However, notable differences arise on the blue side. As expected, the \noigm model closely follows the observed profile, showing a weak but noticeable blue peak at $z<4.0$ that nearly disappears at $z>4.0$. 

In comparison, \igmz and \igm behave similarly. Overall, the \lya stacked line profile predicted by both ANN models remains unchanged through redshift. Both ANN models predict that the red peak becomes slightly narrower with increasing $z$, as found in the observations. However, both \igmz and \igm predict that the amplitude of the blue peak is relatively constant and that it becomes slightly narrower at higher redshift. {  This suggests that the physical mechanisms shaping the \lya line profile  do not evolve strongly through cosmic time. }

In summary, our reconstruction of the pre-IGM spectra using stacks strongly supports the picture suggested by \citet{Hayes_2021} in which the evolution of the observed \lya line profile shape is mainly due to the IGM. This further validates our pre-IGM line profile reconstruction. While they use a similar dataset (obtained from \texttt{LASD}; \citealp{Runnholm_2020}), the method is very different from ours. While they construct a mean IGM transmission from empirical models, we made our analysis on a basis of source per source procedure. The strength of our methodology in comparison to theirs is that \zelda provides the reconstructed spectrum of individual sources affected by very different IGM line of sights.


\subsection{ Comparison between \mfa and the IGM mean transmission}\label{sec:comparison_mfa_IGM}

\begin{figure*} 
        \includegraphics[width=3.6in]{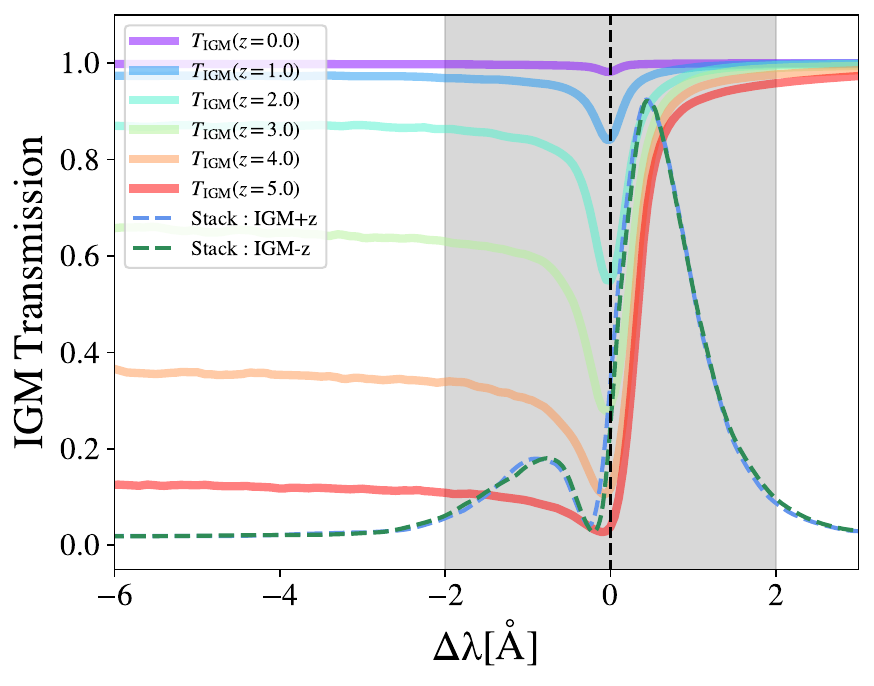}
        \includegraphics[width=3.6in]{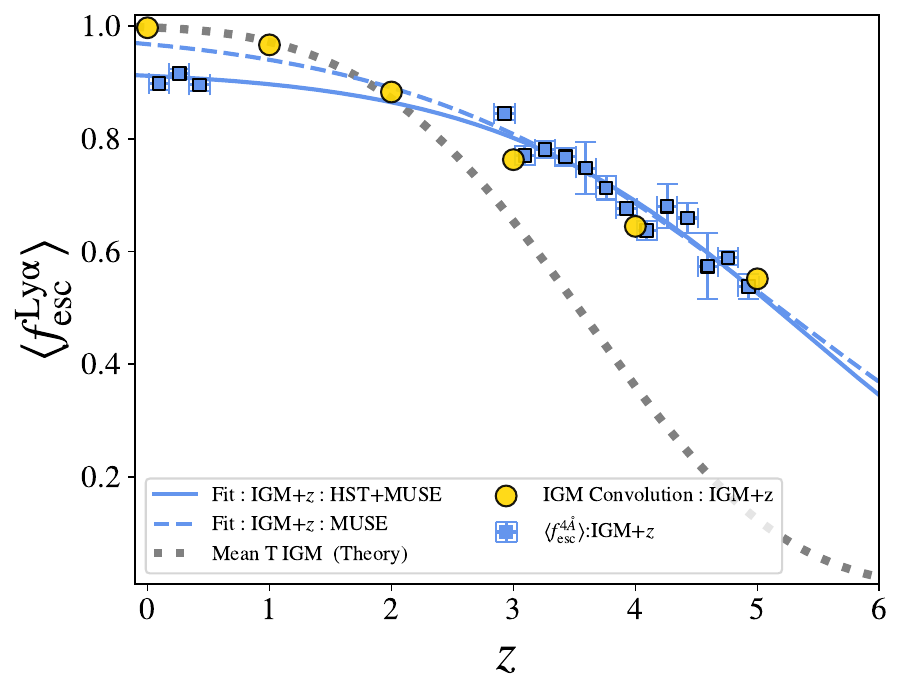}
        \caption{ {\bf Left:} illustration of the Lya emission line IGM escape fraction computation from the stacked line profile. The solid colored lines show the mean IGM transmission of \cite{Byrohl2020} in increasing redshift from purple to red. These curves are recalibrated to match the mean IGM transmission of \cite{Faucher-Giguere08}. The dashed green and blue line show the reconstructed stack line profiles using zELDA models (IGM+z in blue and IGM-z in green). The shaded region marks the wavelength window there the Lya emission line IGM escape fraction is computed.  {\bf Right:}  Mean Lya IGM escape fraction as a function of redshift (squares) for IGM+z (blue). The solid lines indicate the fits, including all the sources. Meanwhile, the colored dashed line indicates the fit using only MUSE data. The grey dashed line shows the \cite{Faucher-Giguere08} mean IGM transmission. The  yellow circles show the mean Lya IGM escape fraction using the \cite{Faucher-Giguere08} recalibrated IGM transmission curves applied to the reconstructed stacked spectrum (i.e., convolving the solid lines with the dashed lines from the left panel). }
        \label{fig:mfa_check}
        \end{figure*}

The connection between the average IGM transmission, \mT, and the mean IGM \lya escape fraction, \mfa, is initially unknown. \fa quantifies the fraction of \lya radiation escaping from LAEs over intermediate IGM scales (few cMpc), after being perturbed by RT in the ISM and CGM. \mT is derived from the spectra of extragalactic sources within a wavelength window located several angstroms blueward of the \lya line, measuring the IGM hundreds of cMpc from the source. As a first approximation, the IGM primarily affects the blue side of the \lya line. Consequently, the measurement of \fa around the \lya wavelength naturally distinguishes \mT from \mfa. In addition, \fa is obtained by convolving the \lya line profile with the IGM transmission along a sightline. Therefore, the shape of the \lya line profile is crucial for assessing \mfa, whereas \mT remains independent of it. For instance, given a constant IGM transmission, a line profile with a stronger red peak than blue peak yields a higher \fa, and vice versa.

To clarify the distinction between these quantities, we define the Ly$\alpha$ IGM escape fraction within a $4\rm\AA$ window as the ratio of the transmitted flux to the intrinsic flux:
\begin{equation}
    f_{\rm esc}^{4\rm\AA} = \frac{\int_{-2\rm\AA}^{+2\rm\AA} F_{\lambda}^{\rm int}(\Delta \lambda_0) \, T_{\rm IGM}(\Delta \lambda_0) \, d\Delta \lambda_0}{\int_{-2\rm\AA}^{+2\rm\AA} F_{\lambda}^{\rm int}(\Delta \lambda_0) \, d\Delta \lambda_0},
\end{equation}
where $F_{\lambda}^{\rm int}$ is the intrinsic Ly$\alpha$ line profile emerging from the ISM/CGM, $T_{\rm IGM}$ is the IGM transmission along the line of sight, and $\Delta \lambda_0$ is the rest-frame wavelength difference from the Ly$\alpha$ resonance. 

In contrast, the mean IGM transmission is defined independently of the emission line shape as:
\begin{equation}
    \langle T_{\rm IGM} \rangle = e^{-\tau_{\rm eff}},
\end{equation}
where $\tau_{\rm eff}$ is the effective optical depth of the IGM measured from the unabsorbed continuum of the Ly$\alpha$ forest.

To compare our measurement of \mfa with previous works, we compute a synthetic estimate of \mfa based on the redshift evolution of \mT from \cite{Faucher-Giguere08}.  Briefly, \cite{Faucher-Giguere08} measured the IGM effective optical depth ($\tau_{\rm eff}$) using the \lya forest in a large sample of high-resolution quasar spectra. By modeling the unabsorbed quasar continuum, they derived an empirical power-law fit for the redshift evolution of the mean transmitted flux. Specifically, we calculate the mean \lya escape fraction, \mfa, that corresponds to the mean IGM absorption reported by \cite{Faucher-Giguere08}, by convolving the mean IGM transmission curves around \lya with the average \lya line profile as a function of redshift. For this, we adopt the IGM transmission shape from \cite{Byrohl2020}, recalibrated to match \cite{Faucher-Giguere08}. For the mean \lya line profile emerging from the ISM/CGM, we use the reconstructed stacked profile discussed in \ref{ssec:results_observations_stack}, which includes all sources from $z=0$ up to $z=5$. We repeated the analysis using stacked profiles in redshift bins and obtained consistent results, as the stacked spectrum varies little with redshift.

In the left panel of Fig.~\ref{fig:mfa_check}, we show the mean IGM transmission curves as a function of redshift (solid colored lines). The reconstructed stacked  spectrum between $z=0$ and $z=5$, as modeled by \igmz and \igm, is displayed in dashed blue and green lines, respectively. We compute the synthetic \mfa as the ratio between the convolution of the mean IGM transmission and the stacked line profile, and the unabsorbed stacked profile, within a window of $\pm 2\AA{}$ around \lya (grey region).

The synthetic \mfa derived from the stacked line profile and mean IGM transmission curves is shown as yellow dots in the right panel of Fig.~\ref{fig:mfa_check}. The synthetic \mfa decreases with redshift because the stacked line profile emerging from the ISM/CGM remains constant, while the blue peak is increasingly absorbed at higher $z$ as \mT decreases (dashed grey curve). We find excellent agreement between the synthetic \mfa and our measurements from the \lya line profiles (blue squares)  at $z>3.0$ , for both the \igmz and \igm (not shown) models. At $z>2$, the synthetic and measured \mfa values are consistent within $1\sigma$, indicating good agreement with \mT. At $z<1$, however, we find that the measured \mfa is $\sim 8\%$ lower than the synthetic \mfa obtained using \cite{Faucher-Giguere08}.

Considering the parametric estimation of the redshift evolution of \mfa, we find that when only MUSE data are used (dashed blue line), there is very good agreement with the synthetic \mfa, showing consistency with the \mT evolution from \cite{Faucher-Giguere08}. However, when including the HST dataset, the parametric estimate lies $\sim 8\%$ below the synthetic \mfa at $z<1.0$, while remaining consistent at $z>2$.

This result indicates that \zelda predicts \lya line profiles to be more absorbed than expected given the mean IGM optical depth at $z<1$, whereas at $z>1$ the predictions match the literature \mT values. In \zp, we tested biases and accuracy in recovering \mfa as a function of redshift, showing that both \igmz and \igm correctly recover \mfa at low redshift for a range of \mfa$(z)$ relations. Therefore, our findings may indicate either a limitation of the shell model at low $z$, or that \lya line profiles are indeed more attenuated than expected at $z<0.5$. In Sect.~\ref{sec:disucssion_igmz_at_low_z}, we show several examples of $z<0.5$ sources that may exhibit IGM-attenuated \lya spectra, supporting this interpretation.

For comparison, we also show \mT, the mean IGM transmission from \citep{Faucher-Giguere08}, as a dashed grey line. At $z<0.5$, the \igm and \igmz predictions of \mfa fall $\sim 0.1$ below \mT. The difference between \fa and \mT can be attributed to stronger absorption around the \lya line center \citep{laursen11,gurung_lopez_2022}. Even if the IGM transmission is close to unity a few angstroms away from \lya, stronger absorption in the range $-2\AA{}<\Delta \lambda <0.5\AA{}$ can affect the blue peak of $z=0$ lines. At higher redshift ($z>2.0$), both the best-fit parametric curves for the ANN models and the predicted \mfa lie above \mT. This is because most of the \lya flux resides in the red peak of the \lya line profile, which remains largely unaffected by the IGM up to $z=5.0$. Thus, although \mT can be as low as $\sim 10\%$ at $z=5.0$, a large fraction of \lya photons still escape through the IGM, as they are already redshifted relative to the \lya wavelength by the ISM/CGM. However, this trend could also reflect a selection effect: in this work, we measure \fa in MUSE galaxies with strong observed \lya emission rather than in the full galaxy population. Consequently, flux-limited samples may yield \mfa values higher than those of the overall star-forming galaxy population.  Quantifying this effect is non-trivial due to the complex dependencies between intrinsic line morphology, luminosity, and detection completeness. For instance, profiles dominated by red peaks suffer less attenuation than double-peaked profiles, creating a selection function that depends on spectral shape. However, the observed scatter of $f_{\rm esc}^{\rm IGM}$ around the median trend is relatively symmetric, suggesting that the sample is not exclusively skewed towards high-transmission sightlines but captures a representative range of attenuation scenarios.

\subsection{Comparison between the \lya global and IGM escape fractions}\label{sec:disucssion_f_esc_global}

\begin{figure} 
        \includegraphics[width=3.4in]{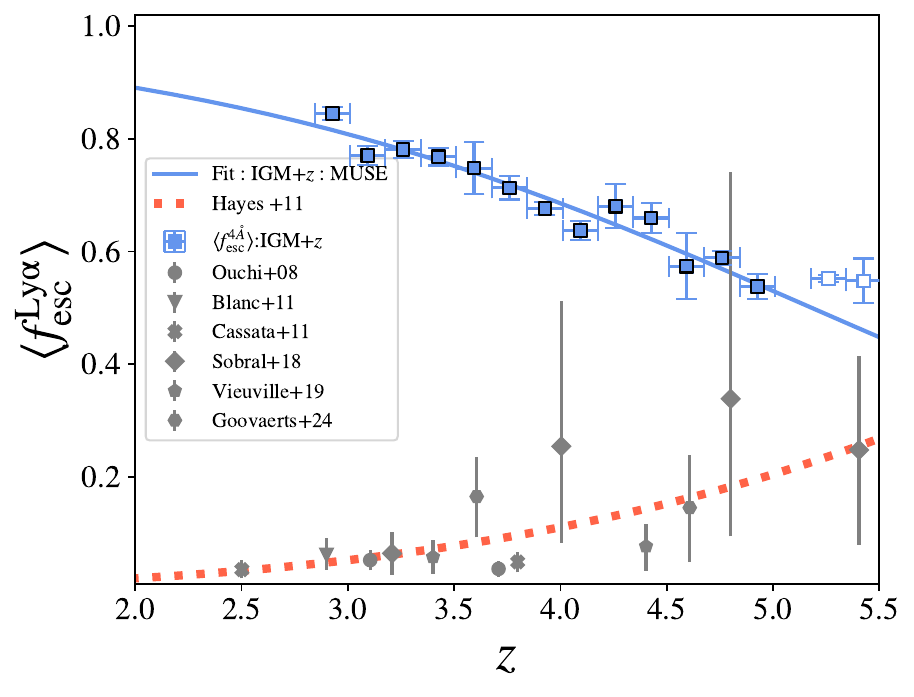}%
        \caption{  Comparison between the redshift evolution of the IGM-only and global \lya escape fractions. The filled colored squares show the IGM \lya escape fraction (\mfa) found by \igmz up to $z=5.0$, while those empty show the results at $z>5$. The colored solid line show the best fit to \mfa as presented previously. The dashed red line shows the global \lya escape fraction estimation by \cite{Hayes11}. The grey symbols show different studies from the literature measuring the global \lya escape fraction and their uncertainty. 
 }
        \label{fig:f_esc_global}
        \end{figure}

\begin{figure*} 
        \includegraphics[width=7.2in]{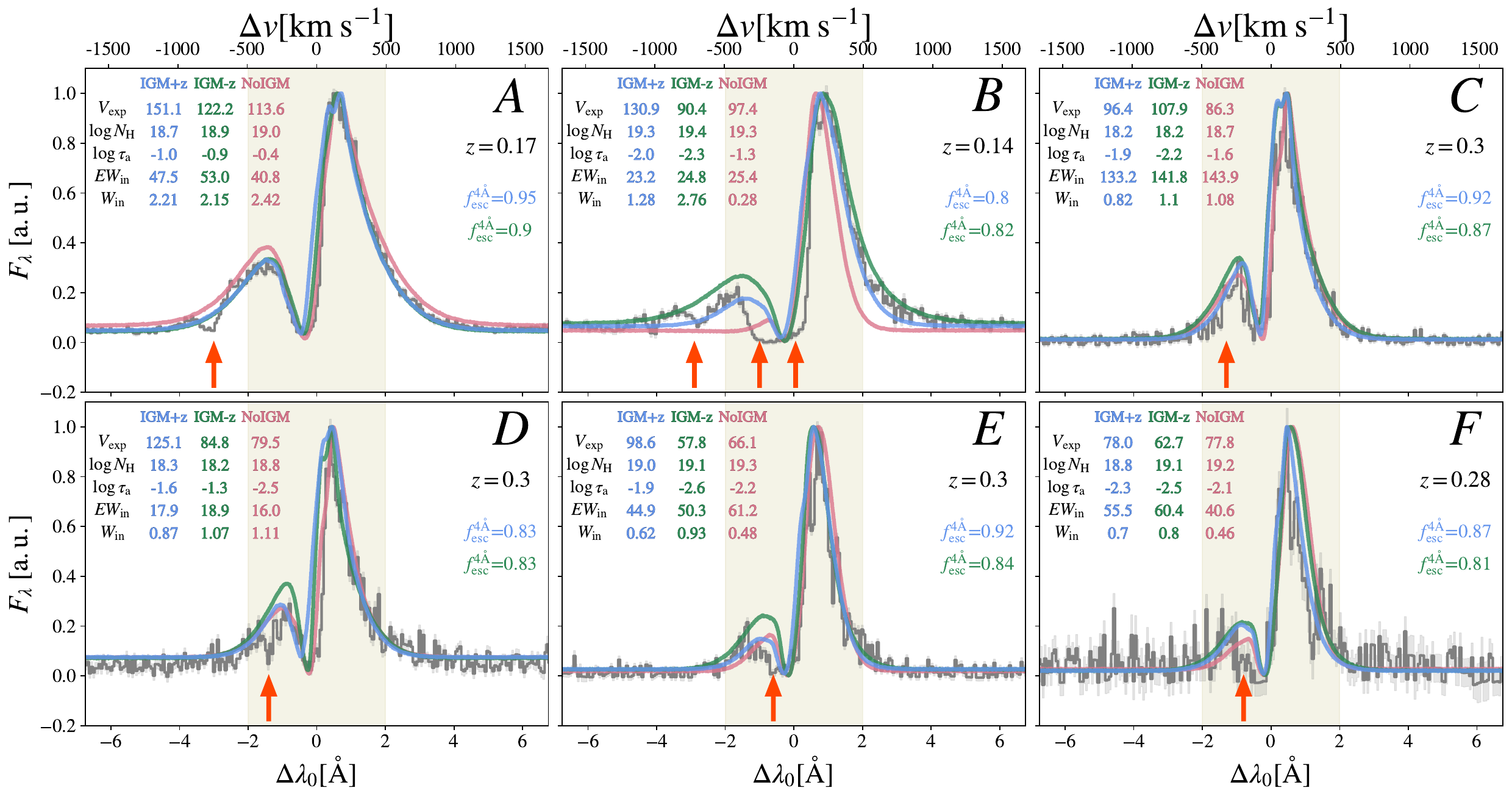}
        \caption{ Six examples of \zelda's prediction on observed line profiles displayed in rest frame. The observed line profile is displayed in dark grey with its $1\sigma$ uncertainty in light grey. \zelda's galactic line profiles of \igmz, \igm and \noigm are displayed in blue, green and pink, respectively. The redshift of the source and the outflow properties for each model are shown in the corresponding color. \vexp is given in \kms, \nh in $cm^{-2}$ and \ew and \w in \AA{}.The central grey shaded region shows the wavelength interval of \fa  . The red arrows mark potential IGM features. }
        \label{fig:LASD_examples_IGM}
        \end{figure*}

To our knowledge, this is the first time that the mean \lya emission-line IGM escape fraction has been directly measured. Consequently, we lack independent measurements from other studies to validate our results. However, the global Ly$\alpha$ escape fraction has been investigated in the literature \citep{blanc11,ouchi08,Cassata_2011,Sobral_2018a,Vieuville_2019,Goovaerts_2024}. In particular, \cite{Hayes11} estimated the redshift evolution of $f_{\rm esc}^{\rm Ly\alpha}$ by comparing the ${\rm H}\alpha$ flux—largely unaffected by radiative transfer—to the measured \lya flux in the same sources. The global \lya escape fraction, $f_{\rm esc}^{\rm Ly\alpha}$, is defined as the ratio between the total Ly$\alpha$ photons produced inside a galaxy and those that reach the observer. It therefore encapsulates the combined effects of the ISM, CGM, and IGM. Since $f_{\rm esc}^{\rm Ly\alpha}$ must be lower than the individual escape fractions of these components, it can serve as a useful validation tool for our measurements.

In Fig.~\ref{fig:f_esc_global}, we show the $f_{\rm esc}^{\rm Lya}$ redshift evolution obtained by \cite{Hayes11} (dashed red line) along with the individual measurements from the other studies in grey as described in the legend. Other studies, such as \cite{Goovaerts_2024} (based on \cite{Thai_2023}), also obtained relations between the redshift and $f_{\rm esc}^{\rm Lya}$ , finding compatible results with \cite{Hayes11}. Overall, it is found in the literature that $f_{\rm esc}^{\rm Lya}$ increases with redshift, being earlier galaxies less attenuated in \lya. Normally, this is attributed to the dust in the ISM of galaxies, which is less abundant at high redshift \citep{Hayes11,GurungLopez_2019a,Goovaerts_2024}. In particular, the different experiments find that $f_{\rm esc}^{\rm Lya}<5\%$ at $z<3.5$. Also, at $5<z<6$, \cite{ouchi08,Cassata_2011,Vieuville_2019} found a $f_{\rm esc}^{\rm Lya}$ below 0.2 with relatively low uncertainty of $\sim 0.1$. Meanwhile, with larger uncertainties, \cite{Sobral_2018a} found a $f_{\rm esc}^{\rm Lya}\sim 0.3$ at $4<z<6$ and \cite{Goovaerts_2024} found  $f_{\rm esc}^{\rm Lya}\sim 0.8$ at $z=6$. Given the uncertainties, all these studies are mutually compatible. 

In this work, we have inferred the IGM escape fraction, \fa, from the observed shapes of Ly$\alpha$ line profiles. By definition, the global Ly$\alpha$ escape fraction, $f_{\rm esc}^{\rm Ly\alpha}$, must always be lower than the individual escape fractions through the ISM, CGM, and IGM, since it is the product of these components. Consistently, our direct measurements of \mfa (blue squares for \igmz) lie systematically above the measured global $f_{\rm esc}^{\rm Ly\alpha}$ (grey dots), which provides a reassuring validation of our methodology.

Moreover, we find that the IGM contributes only marginally to the global \lya escape fraction up to $z \sim 3$, where \mfa$ \sim 0.8$ while the total $f_{\rm esc}^{\rm Ly\alpha}$ remains below 0.1. At higher redshift, as the IGM becomes increasingly dense, its role in regulating the global escape fraction becomes more significant. In particular, at $z=5.0$ \zelda predicts \mfa $\sim 0.5$, while $f_{\rm esc}^{\rm Ly\alpha} \sim 0.2$. This implies that  at this epoch the ISM/CGM absorbs about 40\% of the \lya radiation.  Thus, our results indicate that at redshift 5.0 the IGM is already the main absorber of the \lya emission line. 

Our results therefore suggest that the IGM escape fraction decreases steadily with redshift, at least up to $z=5$. If this trend continues beyond $z=5$, the IGM would become more dominant, controlling the global \lya escape fraction, while the relative impact of the ISM/CGM would diminish.

\subsection{ IGM absorption in \lya line profiles at $z<0.5$}\label{sec:disucssion_igmz_at_low_z}

{  Interestingly, \zelda predicts that the \lya line profiles at $z<0.5$ are already attenuated by the IGM, while the mean IGM transmission is close to unity at $z=0$}. In sections \ref{sec:methodoogy} and \ref{sec:results_observations}, we have discussed the \igmz and \igm models. These two models have very different training sets. On the one hand, \igmz makes use of the IGM transmission curves from \cite{Byrohl2020} without the recalibration\footnote{We refer here to the recalibration to the IGM transmission curves from \cite{Byrohl2020} so that they match the mean IGM redshift evolution by \cite{Faucher-Giguere08}. See \zp.} matching the redshift of the source. On the other hand, \igm uses the IGM transmission curves after recalibration to match the mean IGM transmission in \cite{Faucher-Giguere08}, and the IGM redshift is randomized. These differences cause the \fa distribution to evolve differently with redshift in the training sets (see section 3 in \zp). Thus, the \fa predictions are different between these models. For example, the \igm model gives an unbiased \fa measurement from $z=0$ to $z=5.0$ for mocks with different \mfa redshift evolutions with \sn=15.0 and \w=0.1\AA{}. Meanwhile, for the same spectral quality, \igm overpredicts \mfa by a  5\% when \mfa$\sim 0.8$ at $z<1.0$, while at $z>1.0$ \mfa is recovered correctly (see section 4.3 in \zp).

Despite the differences between \igmz and \igm, both models predict that the \lya line profiles at $z<0.5$ are attenuated by the IGM. \igmz predicts a mean attenuation of $\sim0.92$. Also, \igm predicts \mfa$\sim0.89$. Both models predict individual values ranging from $\sim$0.8 to 1.0. If these lines are IGM attenuated, given the high signal-to-noise and resolution of HST \lya line profiles, a few individual absorbed systems should be visible, {  similar to the \lya forest.}

In Fig.~\ref{fig:LASD_examples_IGM}, we show six examples of \lya line profiles that could be showing IGM features at $z<0.5$. Example {\it A} is a double peak \lya line profile that has a very good signal-to-noise in individual pixels. At $\Delta\lambda_0\sim-3\AA{}$, in the tail of the blue peak, there is an apparent absorption. In contrast, \igmz, \igm, and \noigm predict that the tail of the blue peak decreases smoothly. This kind of feature is not present in the shell model. Thus, this might be an IGM absorption. Moreover, in example {\it A}, the red tail is quite well fitted by the three models. However, the red peak observed spectrum falls faster than the models. This could indicate that the IGM is absorbing the core of the line. This is also visible in examples {\it C} and {\it D}.

Also, in example {\it B} we find a plateau of flux $\sim0$ at the core of the line (from $\Delta\lambda_0=\sim -1.5\AA{}$ to 0.3\AA{}). Note that \zelda has problems reproducing these wide absorption troughs at line center since we restrict ourselves to $T=10^4\,$K gas. Higher (effective) temperatures lead to a wider absorption and thus can alleviate the need for central IGM absorption. This is not the case for absorption features off-line-center also marked in Fig.~\ref{fig:LASD_examples_IGM} . In comparison, the continuum of this source is detected around $0.05$. Additionally, this line shows a wide red peak that extends up to  $\Delta\lambda_0\sim4\AA{}$. Symmetrically, the blue peak seems to extend down to $\Delta\lambda_0\sim-4\AA{}$. As in case {\it A}, example {\it B} shows a decrease in flux in the blue peak at $\Delta\lambda_0\sim-3\AA{}$. Neither of these two features is found in the shell models and might be compatible with IGM absorption. 

Furthermore, in cases {\it C, D} and {\it F}, we find that the blue peak exhibits small drops in flux. However, considering the signal-to-noise of individual pixels, these drops are significant. The three models \igmz, \igm, and \noigm match well the red peak. The models and observations match the ascension of the blue peak ($\Delta\lambda_0$ between -0.5\AA{} and 0.0\AA{}) well -- especially in cases {\it C} and {\it D}. Also, the predicted model is slightly above the observations, while they match again at the tail of the blue peak ($\Delta\lambda_0$ between -2.0\AA{} and -1.8\AA{}). 

Finally, example {\it E} is a double peak \lya line profile with very good signal-to-noise in individual pixels. We find that \igmz and \noigm models predict almost the same intrinsic line profile, while \igmz predicts a line with the same red peak but with a more substantial blue peak. Focusing on the blue peak, we find that in the observed \lya line profile, there is more flux in the tail of the blue peak than in the \igmz and \noigm models. However, the blue tail is very well fitted by the \igm model. An absorbing system in the IGM could also cause this. 

In summary, across examples A–F, potential signatures of IGM absorption are consistently observed in the blue peak, including small flux drops and deviations from the smooth profiles predicted by our shell models. 
{  While more sophisticated intragalactic \lya radiative transfer models might provide somewhat better fits without IGM attenuation, note that generally such models have similar problems in carving out steep absorption features due to the typical diffusive frequency redistribution smoothing spectra. Additionally, features such as steeper declines in the red peaks and wider absorption plateaus at line center could further hint at the impact of IGM on \lya line profiles. Hence, these findings highlight the IGM's potential role in shaping \lya profiles down to low redshifts, particularly in regions where intragalactic radiative transfer models struggle to fully replicate the observed spectra.
}

\section{Summary and conclusions}\label{sec:conclusions}

The observed \lya line profile is the result of the combination of the radiative transfer processes taking place in the interstellar, circumgalactic, and intergalactic mediums. In this work, we have disentangled the IGM from the ISM/CGM components in 313 observed \lya line profiles by using the python package \zelda. 

The latest version of \zelda includes different artificial neural network models
to recover the galactic component from an IGM attenuated \lya line profile. These models are trained using mock \lya lines computed using a grid of precomputed 'shell model' \citep{gurung_lopez_2022} with a full Monte Carlo radiative transfer treatment \citep[\lyart , ][]{orsi12} and the \lya IGM transmission published by \cite{Byrohl2020} based on TNG100 simulation \citep{Nelson_2019}.

We have applied \zelda to 313 observed \lya line profiles, extracted from \lasd \citep{Runnholm_2020}. Of them, 111 come from HST/COS observations, and 202 from MUSE observations. \zelda's results on observed \lya line profile are:

\begin{itemize}
    \item In general, we find that \lya line profiles are little attenuated in HST data (z$<$0.5). Meanwhile, in MUSE data ($z>3.0$), both, \igmz and \igm, predict that the blue side of \lya is attenuated.  The intrinsic ISM Ly$\alpha$ line profiles predicted by our models (i.e., prior to IGM attenuation) frequently exhibit a stronger blue peak than the final observed spectra (which include IGM absorption). This trend becomes prominent at higher redshifts. This suggests that \lya line profile reconstruction becomes more important at higher redshifts. Therefore, galaxy kinematic studies using the \lya line profile at high redshift might benefit from using the IGM reconstructed line profile rather than the observed line profile directly.  \\

    \item We quantify the {  \lya line profile} IGM escape fraction, \fa, for the HST and MUSE samples. We find that the mean \fa at $z<0.5$ is different depending on the ANN model. While \igmz predicts \mfa=0.92$\pm0.02$, \igm predicts \mfa=0.89$\pm0.02$. As discussed, \igmz tends to be biased towards 1.0 in this redshift range, so the measurement provided by \igm should be more reliable. Moreover, both ANN models agree well when determining \mfa in MUSE data. \igmz and \igm predict that \mfa evolve from 0.87$\pm0.04$ at $z=3.0$ to 0.55$\pm0.05$ at $z=5.0$. This measurement can be helpful to obtain a measured 'IGM-free' \lya luminosity function. Interestingly, our models predict that $z<0.5$ \lya line profiles can exhibit IGM features such as those of the \lya forest. We have shown six potential examples of this.  \\

    \item The validation analyses confirm that \zelda provides robust reconstructions of \lya line profiles and reliable estimates of the redshift evolution of the IGM escape fraction. The method achieves sub-angstrom restframe wavelength accuracy ($\sim$0.3\AA{}). The reconstructed profiles reproduce the red tail of the emission line with high fidelity and, when stacked, show little intrinsic evolution with redshift, indicating that the observed spectral variations are primarily driven by IGM absorption, in agreement with previous works \citep{Hayes2021}.
    
    \item  We find that our estimate of the \lya IGM escape fraction is consistent with both the mean IGM optical depth and the global \lya escape fraction reported in the literature. Our results suggest that at $z\sim5$ the IGM is the main absorber of the \lya emission line, attenuating $\sim60\%$ of the total flux, while the combined contribution of the CGM and ISM accounts for the remaining $\sim40\%$. If the observed trend of a decreasing \mfa with redshift extends to higher redshifts, the IGM would become even more dominant than the ISM/CGM in determining the observability of \lya emitters.

\end{itemize}

In times to come, \zelda could facilitate remarkable scientific explorations. For instance, with a substantial sample of LAEs and spectra of sufficient quality, \zelda may enable the 3D mapping of the IGM \lya escape fraction. This mapping could permit the analysis of cross-correlations between \fa and LAEs, potentially shedding light on whether the large-scale characteristics of the IGM are linked to \lya observability \citep{zheng11}.

\section*{Acknowledgements}
The authors acknowledge the financial support from the MICIU with funding from the European Union NextGenerationEU and Generalitat Valenciana in the call Programa de Planes Complementarios de I+D+i (PRTR 2022) Project (VAL-JPAS), reference ASFAE/2022/025.
This work is part of the research Project PID2023-149420NB-I00 funded by MICIU/AEI/10.13039/501100011033 and by ERDF/EU.
This work is also supported by the project of excellence PROMETEO CIPROM/2023/21 of the Conselleria de Educación, Universidades y Empleo (Generalitat Valenciana).
MG thanks the Max Planck Society for support through the Max Planck Research Group, and the European Union for support through ERC-2024-STG 101165038 (ReMMU).
DS acknowledges the support by the Tsinghua Shui Mu Scholarship, funding of the National Key R\&D Program of China (grant no. 2023YFA1605600), the science research grants from the China Manned Space Project with no. CMS-CSST2021-A05, and the Tsinghua University Initiative Scientific Research Program (no. 20223080023).

\newpage
\bibliographystyle{aasjournal}
\bibliography{ref}

\appendix

\section{Distribution of outflow parameters for \igmz and \igm}\label{ssec:prop_1d_2d_distributions}
\begin{figure*}  
        \includegraphics[width=7.2in]{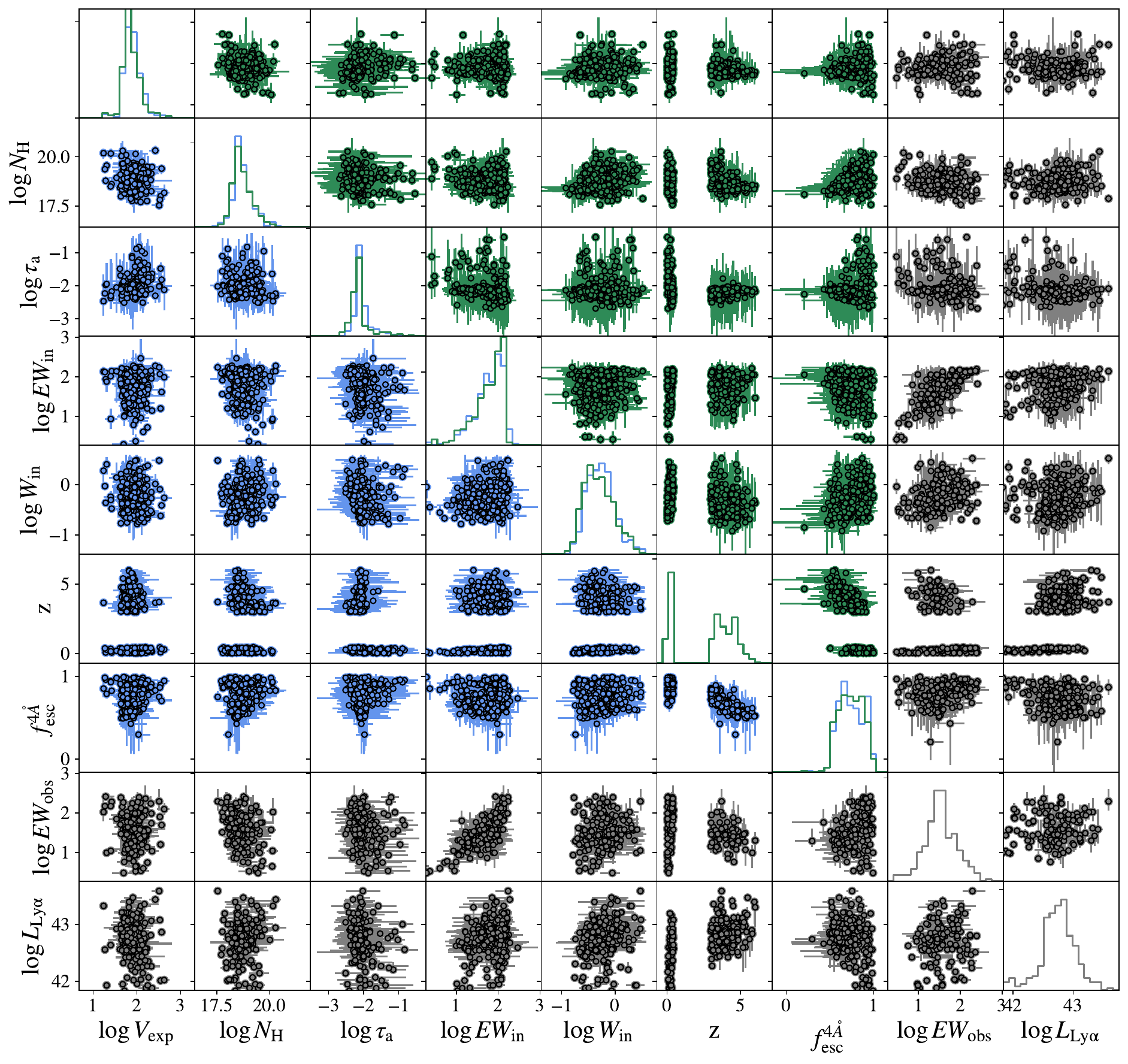}%
        \caption{Comparison of the ANN output and \lasd variables in the observed \lya line profiles predicted by the same models. $EW_{\rm obs}$ and $L_{\rm Ly\alpha}$ were obtained from \lasd. \vexp is given in \kms, \nh in $cm^{-2}$, \ew, $EW_{\rm ob}$ and  \w in \AA{} and $L_{\rm Ly\alpha}$ in $erg/s$. The blue dots show the values predicted by \igmz while the green dots show those of \igm. The grey dots show the properties that include measurements by \lasd. The histograms on the diagonal show the distribution of the parameters in the corresponding color for each model.  The strength of the correlations is quantified in Fig.~\ref{fig:spear_0_7}}.
        \label{fig:all_vs_all}
        \end{figure*}

Fig.~\ref{fig:all_vs_all} shows the predicted parameters plotted against each other: \vexp, \nh, \ta, \ew, \w, $z$, \fa, $EW_{\rm ob}$, and $L_{\rm Ly\alpha}$, arranged from left to right and top to bottom. The upper-right triangle (green and grey) corresponds to the properties predicted by \igm, while the lower-left triangle (blue and grey) corresponds to \igmz. The diagonal panels show the one-dimensional distributions of each parameter.   The strength of the (anti)correlations is quantified in Fig.~\ref{fig:spear_0_7} and discussed in  Sect.\ref{ssec:results_observations_properties}.

\section{ Comparison of Outflow Properties across Models}\label{ssec:results_observations_model_comparison}
\begin{figure*} 
        \includegraphics[width=7.2in]{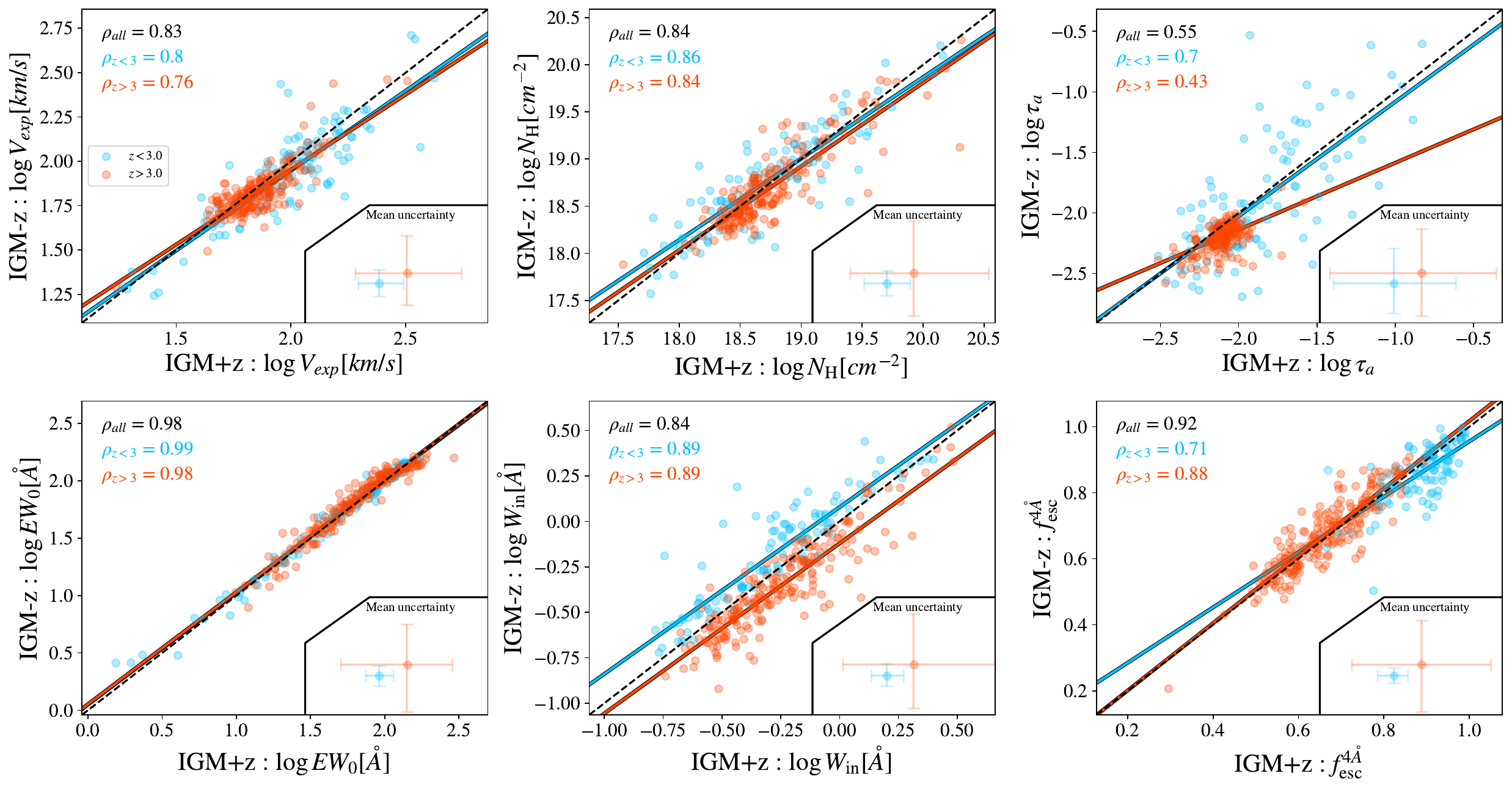}
        \caption{Comparison of the ANN output variables in the observed \lya line profiles predicted by different models. The out given by \igmz is given in the X-axis and \igm in the Y-axis. HST ($z<0.55$) sources are shown in blue, while MUSE ($z>2.9$) sources are shown in red. The black dashed line marks the one-to-one relation. The Spearman coefficient for the full sample is shown in black (top), for HST sources in blue (middle), and for MUSE in red (bottom). The red and blue solid lines are the best-fitting first-degree polynomials for the high and low redshift samples. In the top row \vexp, \nh and \ta are shown from left to right. In the bottom row \ew, \w and \fa are shown from left to right. The mean uncertainty for the shown properties is displayed in the bottom right corner. }
        \label{fig:prop_comp_IGM+z_IGM-z}
        \end{figure*}

\begin{figure*} 
        \includegraphics[width=7.2in]{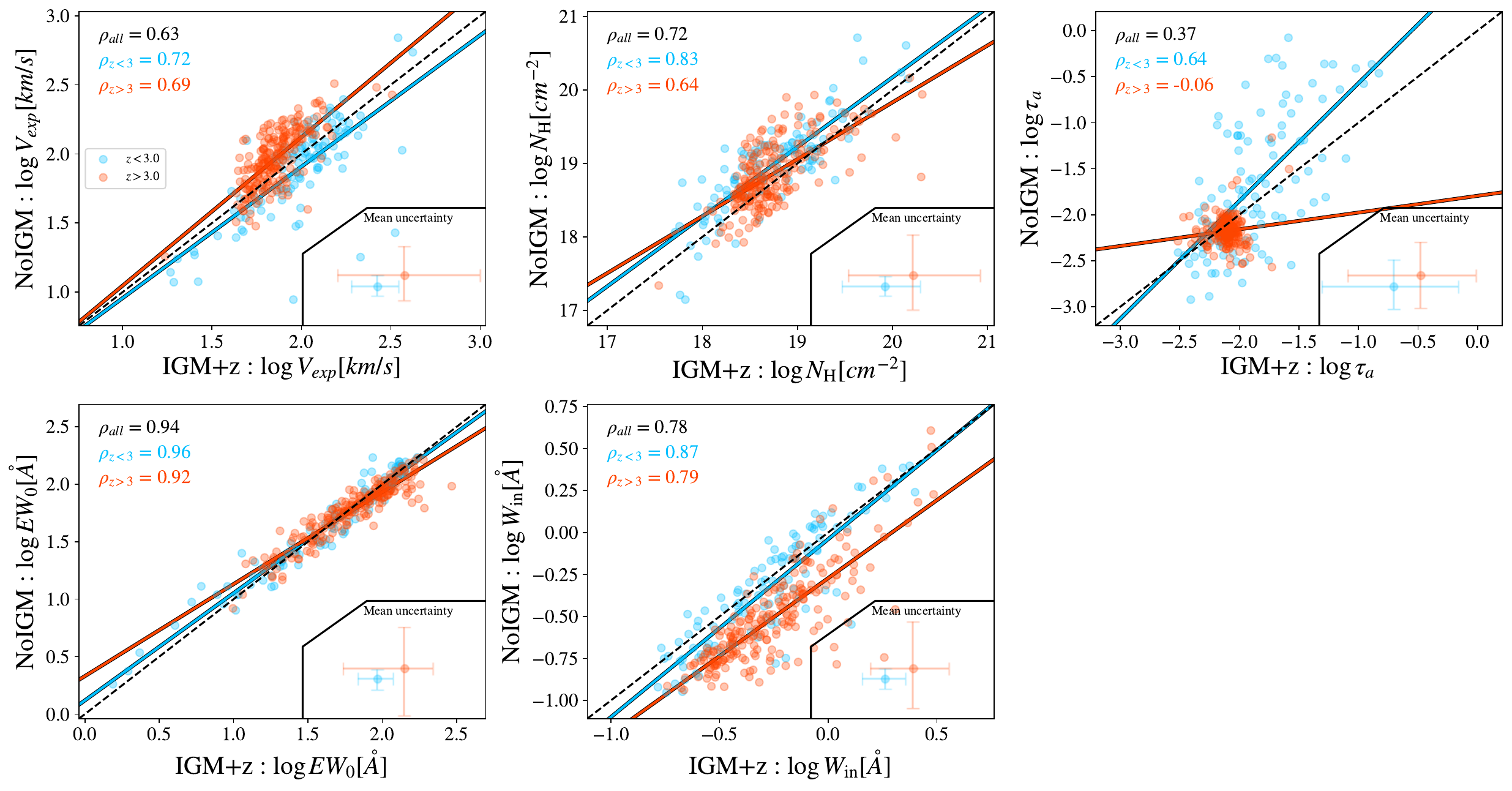}
        \caption{ Same as for Fig.~\ref{fig:prop_comp_IGM+z_IGM-z} but comparing \igmz (X-axis) against \noigm (Y-axis). \noigm does not predict a value for \fa, so we the comparison between \noigm and \igmz for \fa is not available. }
        \label{fig:prop_comp_IGM+z_NoIGM}
        \end{figure*}

In Fig.~\ref{fig:prop_comp_IGM+z_IGM-z}, we compare the outflow properties and \fa predicted by the \igmz (X-axis) and those predicted by \igm (Y-axis). HST sources (low redshift) are shown in blue, while MUSE sources (high redshift) are shown in red. We find that, in general, \igmz and \igm predict the same outflow properties and \fa. The typical Spearman correlation coefficient for the whole sample (shown in the top left corner in black) is above 0.8. For the low and high redshift samples, the output variables also correlate. We find that both models behave almost the same for \vexp, \nh, \ta, and \ew at high and low redshifts. Meanwhile, at high redshift, the \w and \fa values predicted by \igmz and \igm are close to the one-to-one relation. Also, in the low redshift sample, for \w, although the Spearman correlation coefficient is high (0.82), the \w values predicted by \igm are $\sim 0.25\AA$ above the \igmz predictions. Finally, for \fa the low redshift Spearman correlation coefficient is also high (0.71), but \igm predicts slightly lower values than \igmz. 

In Fig.~\ref{fig:prop_comp_IGM+z_NoIGM}, we show a comparison between the output variables of \igmz (X-axis) and those predicted by \noigm (Y-axis). In general, we find that there is a Spearman correlation coefficient higher than $\sim 0.7$ for the whole sample. The Spearman correlation coefficient is greater in the low redshift sample than in the high redshift sample. For example, for \vexp, $\rho_{z<3}=0.72$ while $\rho_{z>3}=0.69$ and for \nh $\rho_{z<3}=0.83$ while $\rho_{z>3}=0.64$. The \igmz model predicts higher values for \vexp at high redshift. Meanwhile, \igmz predicts lower values of \w than \noigm at high redshift. 

The results in Fig.~\ref{fig:prop_comp_IGM+z_NoIGM} reveal systematic biases in shell model fits that do not account for IGM absorption \citep[e.g.][]{Gronke2017}. These models overestimate expansion velocities by $\sim 0.30$\,dex and neutral hydrogen column densities by $\sim0.25$dex. This occurs because the IGM primarily absorbs the blue side of the line -- hence, when models exclude the IGM, they compensate by increasing both the galactic absorption's blueshift (\vexp) and the strength (\nh).



\section{ Possible biases when fitting an IGM attenuated line with a model without IGM }\label{sec:biaseses}

\begin{figure*} 
        \includegraphics[width=7.2in]{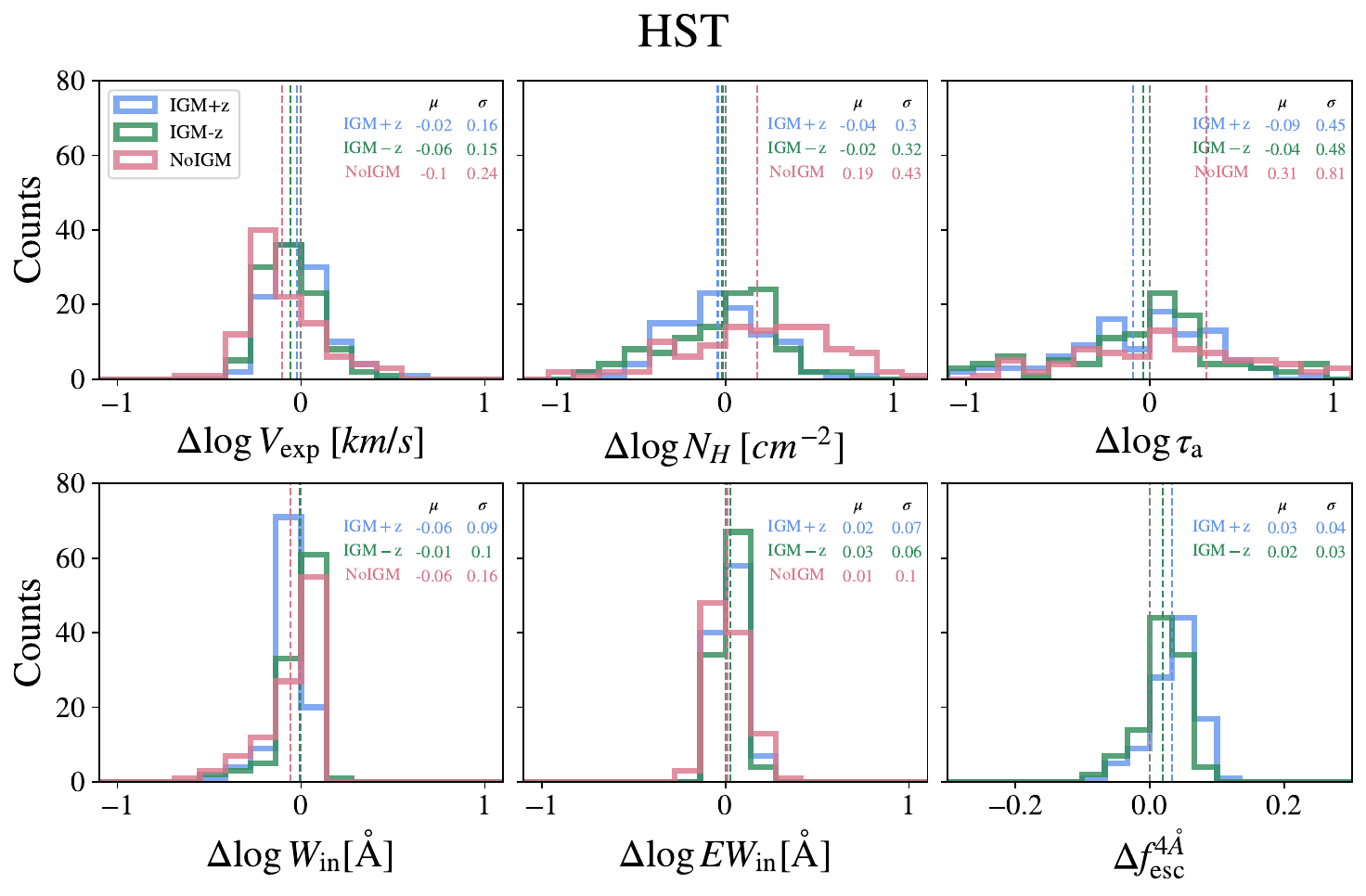}%
        \caption{ Distribution of the difference between the input shell model properties and those recovered for the mock with HST-like quality. In the top panels \vexp, \nh and \ta are shown from left to right. In the bottom panels, \w, \ew, and \fa are shown from left to right. The prediction using \igmz, \igm, and \noigm is shown in blue, green, and pink, respectively. The dashed horizontal lines indicate the distribution's mean, matching the color used in the model. The black dashed line is set at 0 for reference. In top right corner a table summing up the mean ($\mu$) and standard deviation ($\sigma$) is displayed for the three models.    }
        \label{fig:mocks_bias_HST}
\end{figure*}

\begin{figure*} 
        \includegraphics[width=7.2in]{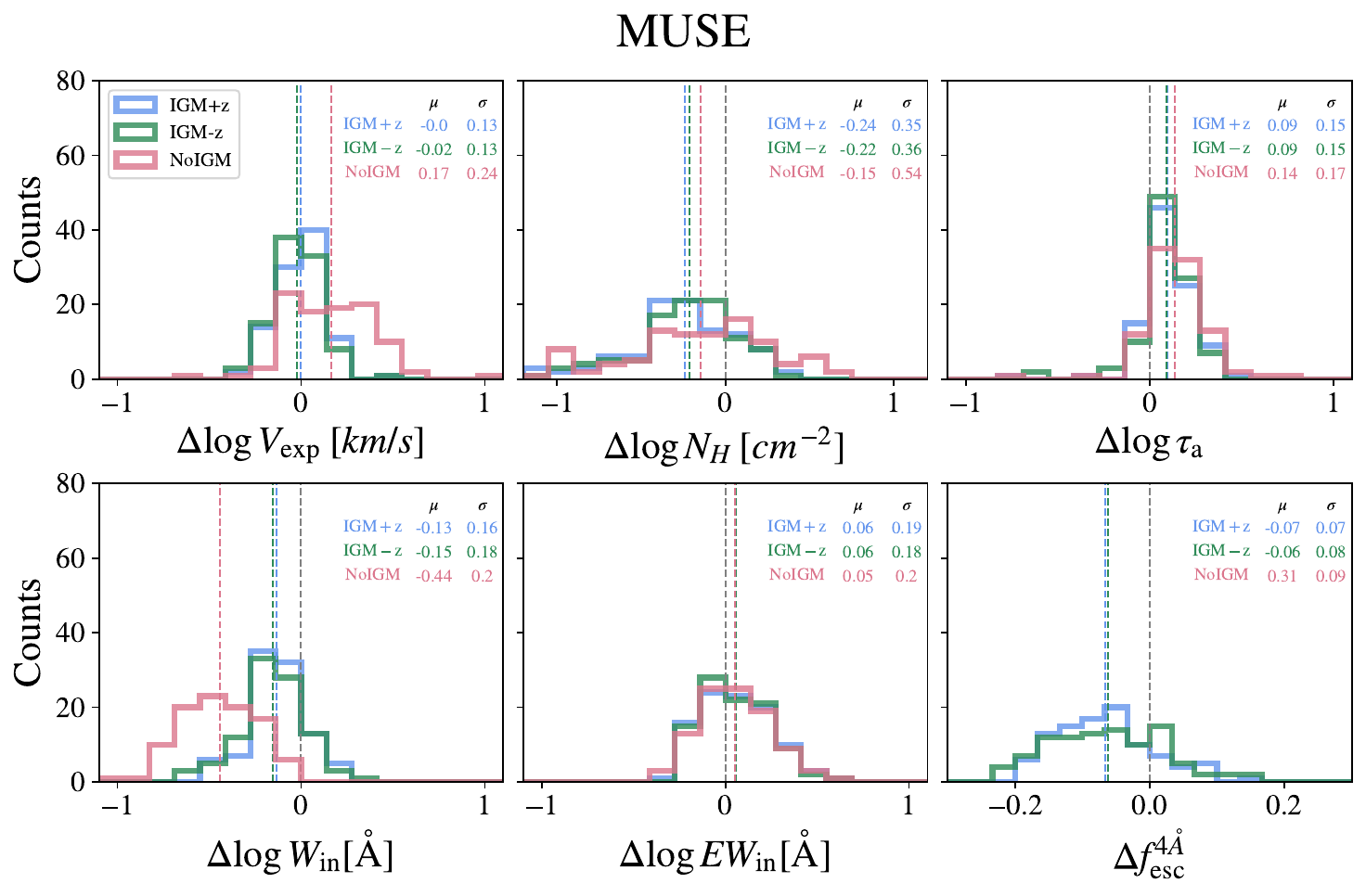}%
        \caption{  Same as Fig.~\ref{fig:mocks_bias_HST}, but for the mocks with the spectral quality of MUSE.
 }
        \label{fig:mocks_bias_MUSE}
        \end{figure*}

The shell model has been widely used to successfully reproduce observed \lya line shapes, providing valuable insights into the physical properties of \lya-emitting systems. However, previous work has often neglected the impact of IGM absorption, which can systematically alter the observed line profiles and bias parameter estimates. In this section, we investigate the possible biases induced by fitting IGM-attenuated \lya line profiles with models that do not account for IGM effects.

In this section, we study the possible biases induced by fitting an IGM-attenuated line with a model without IGM. For this goal, we use \noigm as the model without IGM absorption. Meanwhile, we also compare the results for the explored models with IGM, \igmz, and \igm.   

The bias in the shell model parameters in a given \lya line profile population might depend on the distribution of actual shell model parameters intrinsic to the sample. To make our analysis realistic, we build mock \lya line profile samples from the shell model parameters measured in the observed \lya line profiles in Sect.~\ref{sec:results_observations}. In particular, we produce a single line profile with the same shell model parameters as each of the observed line profiles with matching spectral quality. Then, we convolve the spectrum with an IGM transmission curve at the redshift of the source and measure the \fa. If the measured \fa is within 1\% of that original measure, we keep the IGM transmission curve. Otherwise, we draw another one until the criteria are met. This ensures that the \fa distribution in the mock is very similar to that obtained from the observed \lya line profile. As a result of this procedure, the sample size, redshift distribution, and true shell model parameters in our mock match those of the observed sample. 

In Fig.~\ref{fig:mocks_bias_HST}, we show the distribution of the difference between the measured shell model parameters and the true ones for the mock line profiles of HST. Meanwhile, in Fig.~\ref{fig:mocks_bias_MUSE}, we show those with MUSE spectral quality. In the top right corner, we display the mean and standard deviation of the distributions. Ideally, if the methodology and spectral quality were perfect, the output should match exactly the input, thus all the differences should be null, and the mean and standard deviation should be zero. 

Focusing on the HST-like sample (Fig.~\ref{fig:mocks_bias_HST}), the \igmz and \igm models exhibit biases lower than 0.06 dex, in general. Meanwhile, the \noigm model is unbiased at the level of \igmz and \igm in \w and \ew. In contrast, the mean of the distributions of \noigm is biased towards lower values of \vexp (0.1 dex) and higher values of \nh (0.2 dex) and \ta (0.3 dex). 

Regarding the MUSE-like sample (Fig.~\ref{fig:mocks_bias_MUSE}), the \igmz and \igm models are generally less biased than \noigm. In particular, \igmz and \igm exhibit a bias smaller than 0.1 dex in \vexp, \ta, \ew, and \fa. We find that \igmz and \igm are biased by $0.15$ dex in \w. Moreover, we find that \noigm is more biased than \igmz and \igm, particularly in \vexp and \w. Interestingly, we find that \noigm (0.15 dex) is slightly less biased than \igmz and \igm ($\sim$0.23 dex) when estimating \nh. 

Comparing the  HST-like and MUSE-like samples, we find that the \noigm model is generally more biased than \igmz and \igm. Additionally, we find that the standard deviation of the distribution of differences changes with the spectral quality. In general, we find that the width of the distribution is large in the MUSE-like sample. This direct effect of the lower spectral quality makes the line profile reconstruction more difficult.


\section{ Property correlation in MUSE and HST }\label{app:corr_muse_hst}

\begin{figure*} 
        \begin{center}
        \includegraphics[width=3.6in]{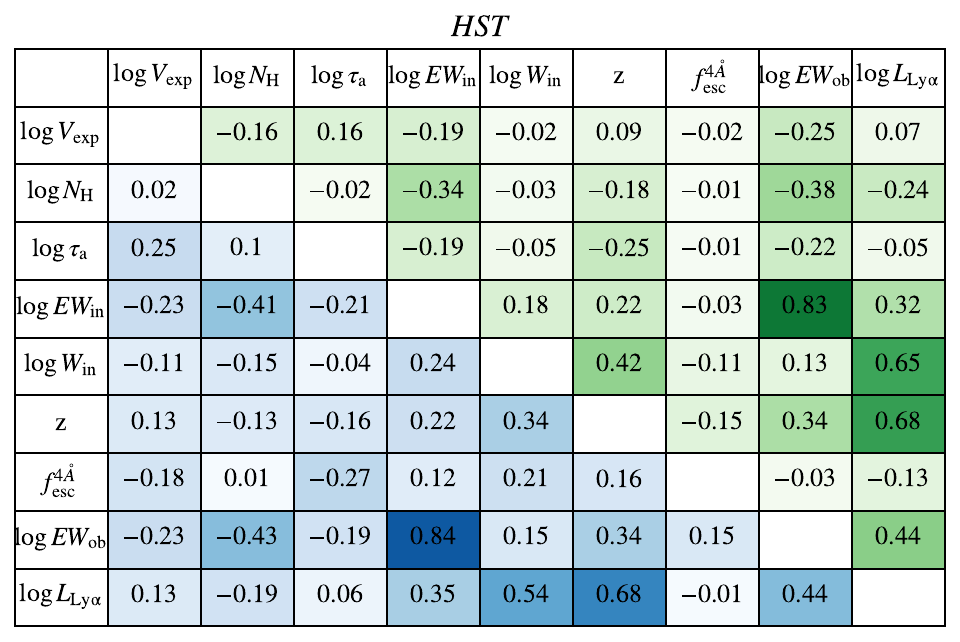}%
        \includegraphics[width=3.6in]{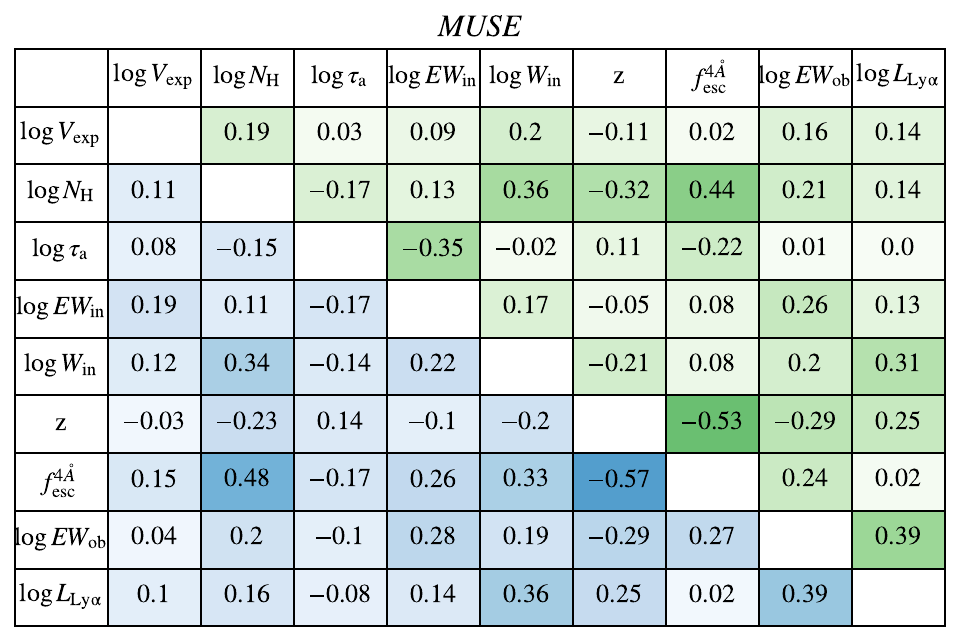}
        \end{center}
        \caption{Same as Fig.~\ref{fig:spear_0_7} but only for HST sources in the left panel and  MUSE sources in the right panel. The uncertainty is not shown for simplicity but is similar to the previous figure's.}
        \label{fig:spear_1_7}
        \end{figure*}
 In Fig.\ref{fig:spear_1_7} we show the Spearman correlation coefficients for the HST (left) and MUSE (right) samples. We find that in general, all the correlations found in the complete sample (see Fig.\ref{fig:spear_0_7}) are also present in the individual samples, such as the correlation between \ew and $EW_{\rm ob}$. 

 Other correlations existing in the combined sample are only present in one of them. For example, the anti-correlation between \fa and redshift is present in the MUSE and combined sample. However, no trend is found within the HST sources. 


\section{  \fa inference by an alternative model}\label{sec:disucssion_other_models}

\begin{figure*} 
        \includegraphics[width=7.2in]{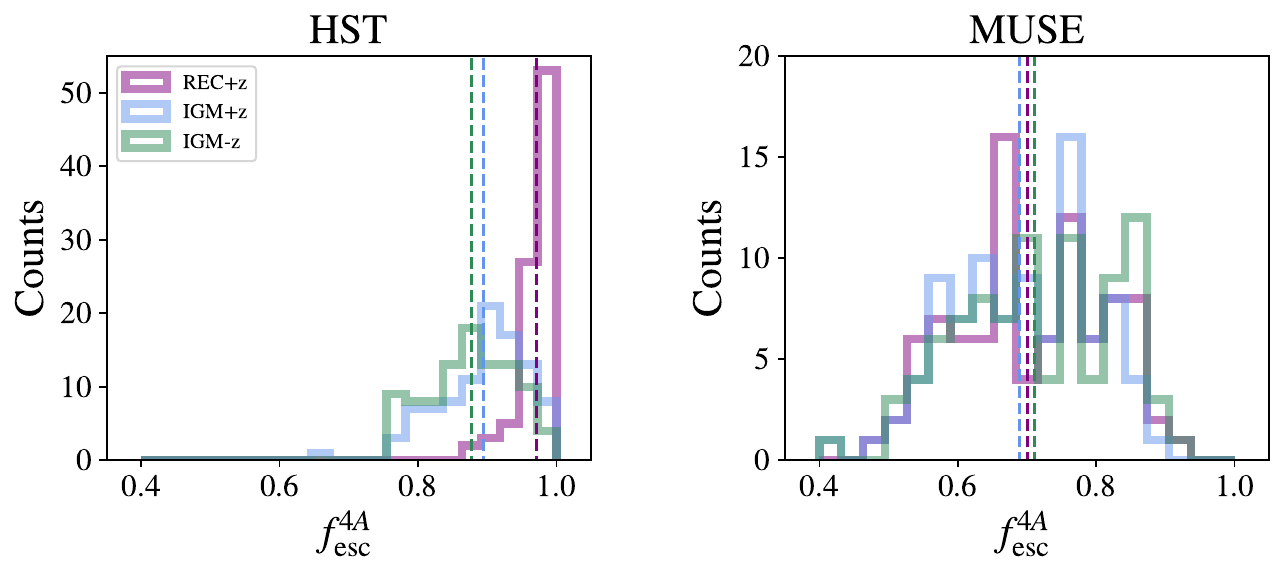}%
        \caption{ Distributions of \fa predicted values for HST  (left) and MUSE \lya line profiles (right). The distribution for \recz, \igmz, and \igm are shown in purple, blue, and green, respectively. The vertical dashed colored lines indicate the distribution median for the corresponding model. }
        \label{fig:rec_f4a}
        \end{figure*}

In this section, we discuss the \fa provided by a \zelda alternative model, \recz, fully introduced in Section 5.2 of \zp. \recz is built to use the recalibrated IGM transmission curves (as \igm), and the redshift of the source is included in the input (as \igmz).  In comparison with \igmz and \igm,  the \recz training set shows a similar \fa distribution as at $z>2.0$; however, at $z<1.0$, the dispersion in \recz is significantly smaller. In particular, 2$\sigma$ of the sources have \fa$>$0.9. In contrast, in the \igmz training set, more than 2$\sigma$ of the sources exhibit \fa$>$0.8 at  $z<1.0$. This is caused by the fact that the IGM absorption at the \lya wavelength is heavily reduced after the recalibration to match \cite{Faucher-Giguere08}. The small scatter causes that \recz \fa measurements are biased towards high values at $z<1.0$.

We find the same trends in the performance of \igmz, \igm, and \recz in the observed spectrum as in the mock spectrum. The distribution of the predicted \fa values of the observed \lya line profiles presented in Sect.~\ref{sec:results_observations} for \recz (purple), \igmz (blue,) and \igm (green) are shown in Fig.~\ref{fig:rec_f4a}. The left panel shows HST sources at $z<0.55$, while the right panel shows MUSE sources at $z>2.9$. For MUSE sources, the three models show very similar \fa distributions with \mfa $\sim 0.7$ (horizontal dashed lines). Not only is the distribution in $2.9<z<6.0$ the same, but also the redshift evolution, although it is not explicitly shown. Meanwhile, considering HST sources, the \igmz and \igm exhibit very similar \fa distribution, with \mfa $\sim 0.89$ and $0.91$ respectively. Meanwhile, \recz predicts higher \fa values, as the \fa distribution peaks at $\sim 1$ with small dispersion. Actually, the minimum \fa predicted by \recz is $\sim0.9$. Consequently, the \mfa for \recz is $\sim 0.98$.    

The training set in an artificial neural network determines the fidelity of its output. In particular, biases and trends inherent to the training set are imprinted on the output. We find that \recz gives a \mfa of $\sim 1$ at $z<0.5$ in contrast to \igmz and \igm that give \mfa$\sim 0.9$. Meanwhile, the three models give very similar predictions at high redshift. Therefore, the low $z$ discrepancy comes from the training set of \recz. The low \fa variance at $z<1$ in \recz causes its predictions to be biased towards \fa=1 at low $z$, rather than being a real prediction.  

\section{ All fitted lines }\label{app:lines}

In this appendix, we show all the fitted lines for both MUSE and HST data. Line profiles are sorted by redshift. Fig.~\ref{fig:LASD_0} shows sources with $0.0<z<0.20$, Fig.~\ref{fig:LASD_1} for sources with $0.20<z<0.35$, Fig.~\ref{fig:LASD_2} for sources with $0.35<z<3.3$ , Fig.~\ref{fig:LASD_3} for sources with $3.3<z<3.7$ , Fig.~\ref{fig:LASD_4} for sources with $3.7<z<4.2$ ,Fig.~\ref{fig:LASD_5} for sources with $4.2<z<4.8$ and Fig.~\ref{fig:LASD_6} shows sources with $4.8<z<6.0$. The observed \lya line profile is shown in grey, while the prediction of \igm, \igmz, and \noigm are shown in blue, green, and pink, respectively. Furthermore, \fa is shown in the top left corner in the same color as the model used, alongside the source redshift. The grey shaded band marks the wavelength region where \fa is estimated.  

In Fig.~\ref{fig:LASD_bad}, we show the \lya line profiles from \lasd that were not considered for the analysis in this work. They removed these line profiles because i) they looked noisy or ii) they exhibited a very steep continuum not modeled in \zelda, or iii) the fits of \igm, \igmz, and \noigm looked unsuccessful.  From the original sample of 349 spectra, 35 (10\%) were excluded. 

In general, at low redshift (Fig.~\ref{fig:LASD_0} and \ref{fig:LASD_1}) \igmz, \igm and \noigm predict similar intrinsic \lya line profiles.  In sources 7, 40, and 111 \igmz and \igm predict almost the same \lya line profile, and \noigm predicts a different one. In sources 1, 14, 30, 38, 66, 82, and 97 \igmz and \igm actually predict a different intrinsic \lya line profile. Also, in general, the \fa predicted by \igmz and \igm are close to unity, with \igm \fa predictions slightly lower than those of \igmz. 

Meanwhile, at higher redshift (Fig.~\ref{fig:LASD_3}, \ref{fig:LASD_4}, \ref{fig:LASD_5}, and \ref{fig:LASD_6}), the intrinsic \lya line profile predicted by \igmz and \igm match, while they diverge from that of \noigm. Although in some cases, the three models match relatively well, such as in sources 116, 119, 142, 162, 186, 220, and  299. Note that this scenario is more frequent at lower redshift, closer to $z=3.0$. Meanwhile, at higher redshifts, closer to $z=6.0$, there is more difference between the prediction and \igmz/\igm and \noigm. Also, in $\sim$80\% of the high redshift sources, the \igmz and \igm predict the presence of a blue peak in the \lya intrinsic spectrum.  
\begin{figure*} 
        \includegraphics[width=6.9in]{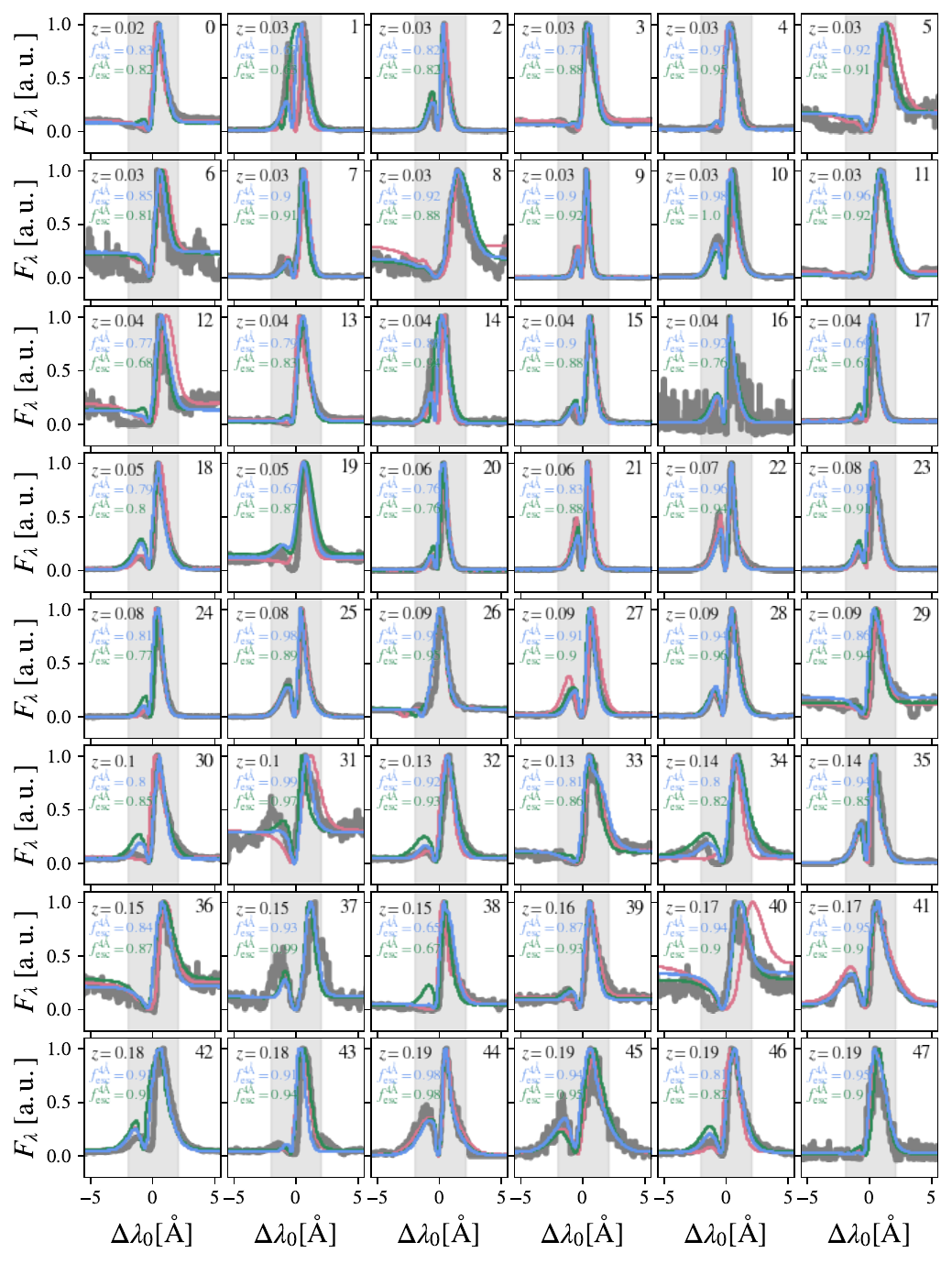}%
        \caption{ \zelda's prediction on observed line profiles displayed in restframe. The observed line profile is shown in grey. \zelda's reconstruction using the models \igmz, \igm, and \noigm is displayed in blue, green, and pink, respectively. The redshift of the source is shown in the top left corner. \zelda's estimations of \fa given by the \igmz and \igm models are shown in blue and green, respectively. The grey-shaded region shows the wavelength interval of \fa. The red number indicates the index in \lasd .  }
        \label{fig:LASD_0}
\end{figure*}
\begin{figure*} 
        \includegraphics[width=7.1in]{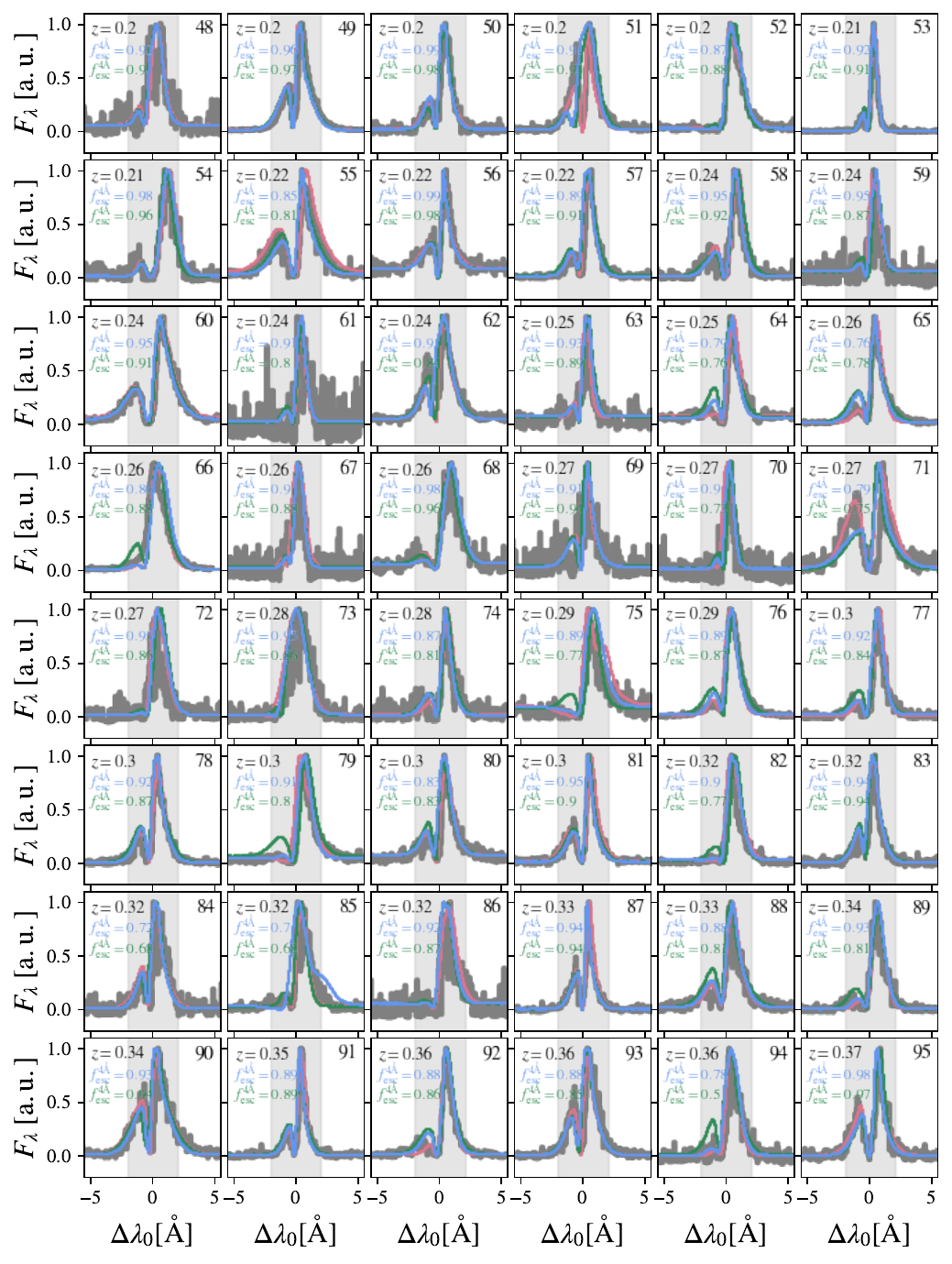}%
        \caption{ Same as Fig.~\ref{fig:LASD_0}, but for other observed \lya line profiles between $z=0.2$ and $z=0.35$.}
        \label{fig:LASD_1}
\end{figure*}
\begin{figure*} 
        \includegraphics[width=7.1in]{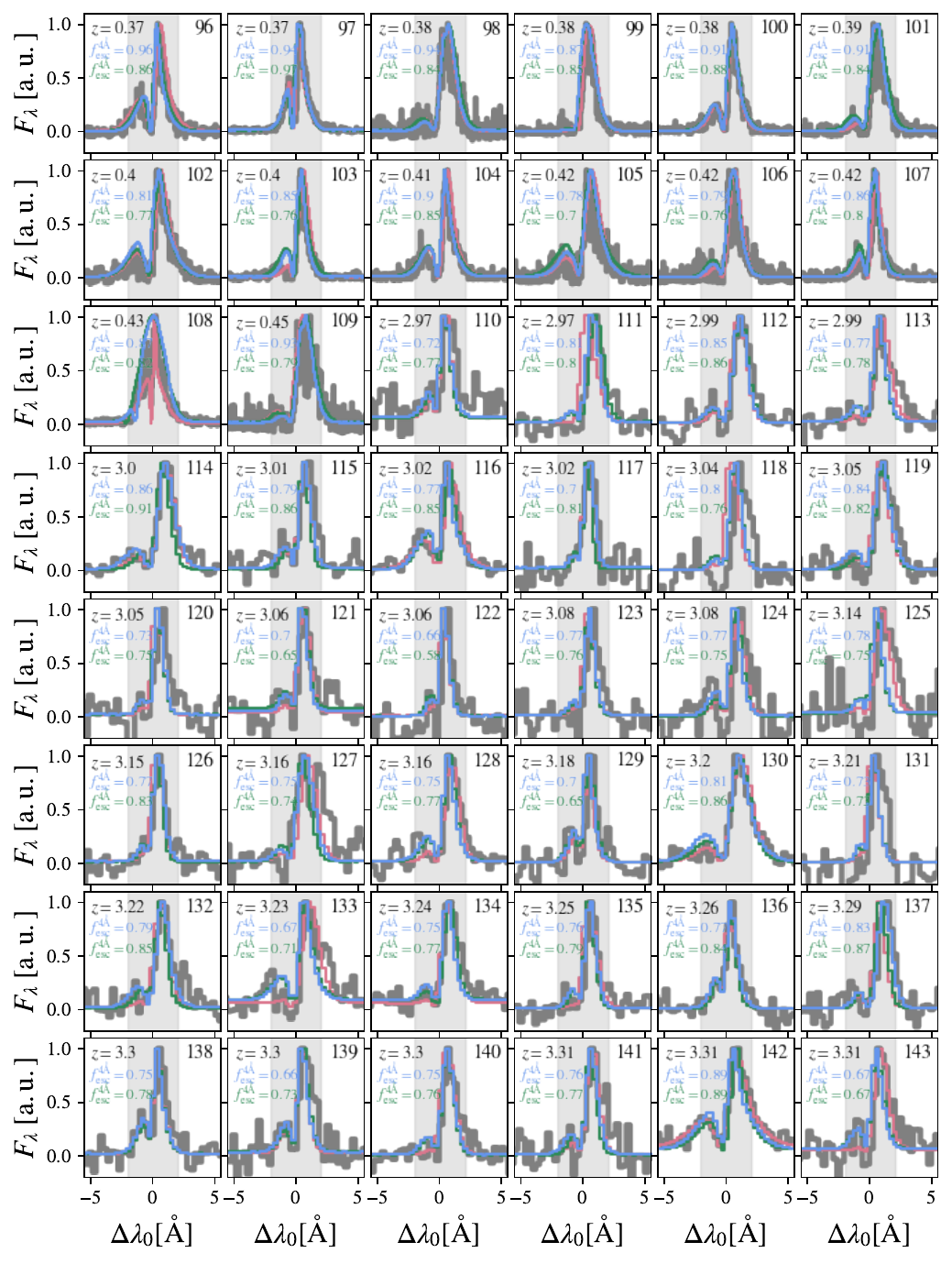}%
        \caption{  Same as Fig.~\ref{fig:LASD_0}, but for other observed \lya line profiles between $z=0.35$ and $z=3.3$. }
        \label{fig:LASD_2}
\end{figure*}
\begin{figure*} 
        \includegraphics[width=7.1in]{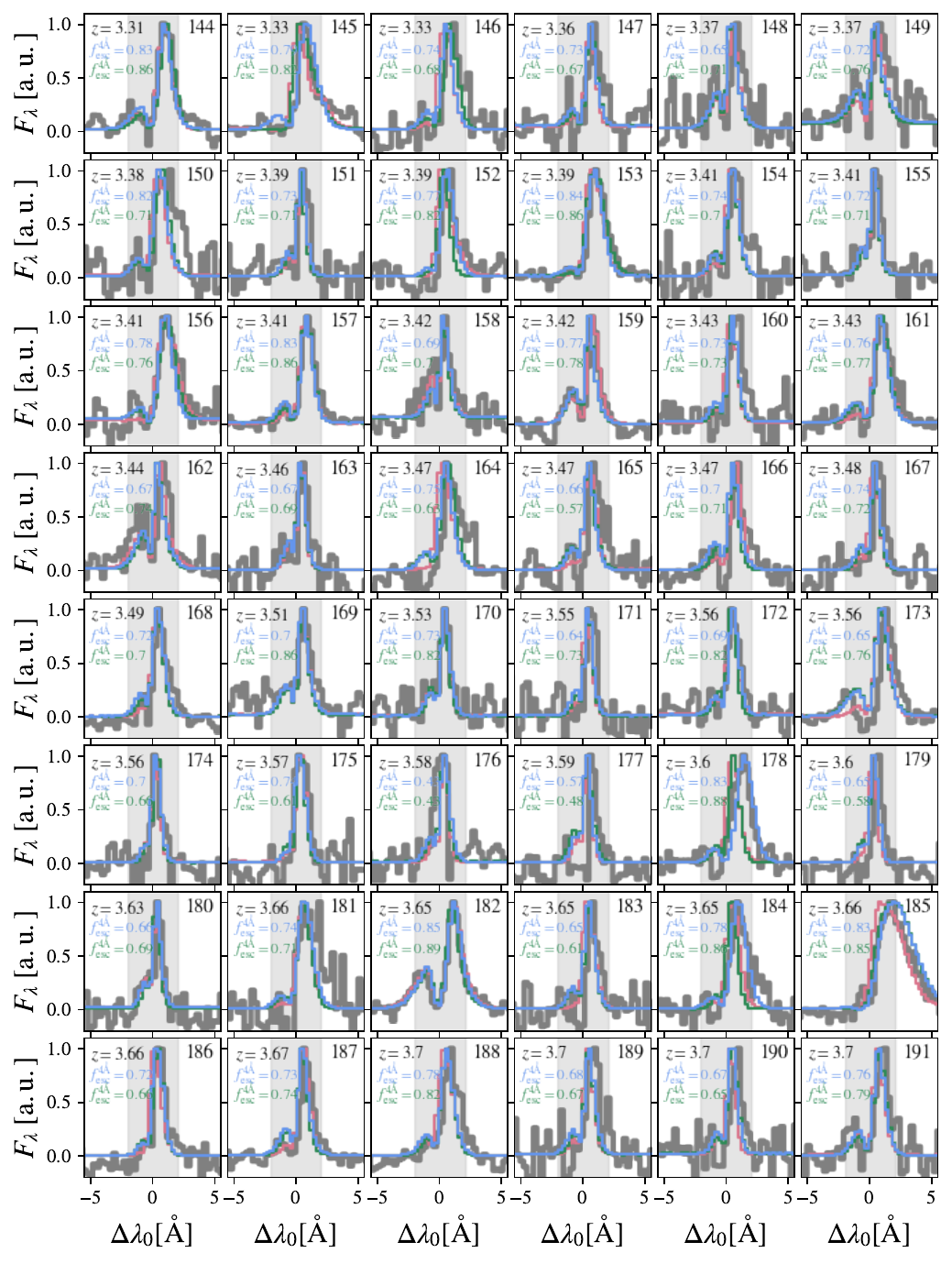}%
        \caption{  Same as Fig.~\ref{fig:LASD_0}, but for other observed \lya line profiles between $z=3.3$ and $z=3.7$. }
        \label{fig:LASD_3}
\end{figure*}
\begin{figure*} 
        \includegraphics[width=7.1in]{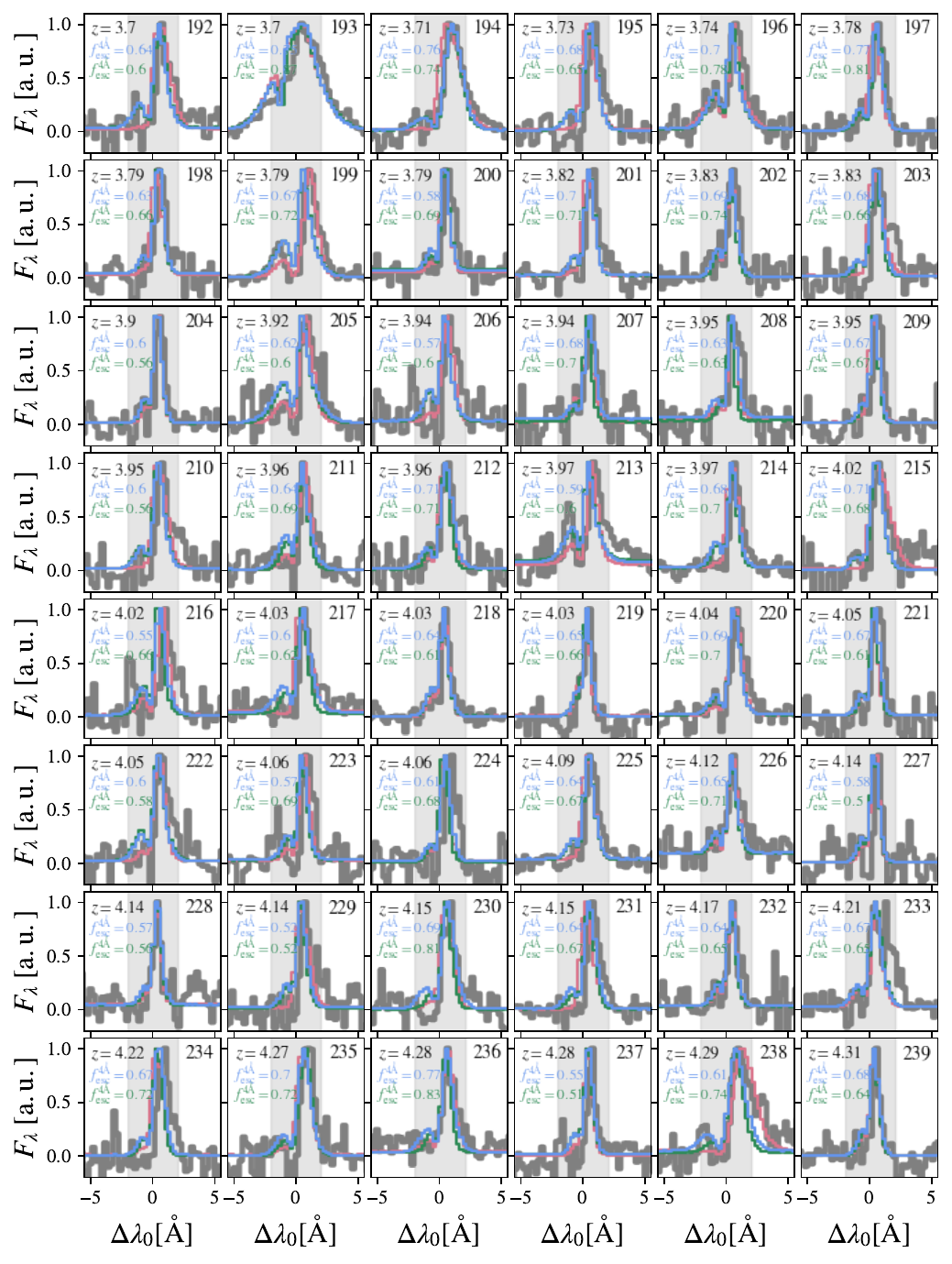}%
        \caption{ Same as Fig.~\ref{fig:LASD_0}, but for other observed \lya line profiles between $z=3.7$ and $z=4.20$. }
        \label{fig:LASD_4}
\end{figure*}
\begin{figure*} 
        \includegraphics[width=7.1in]{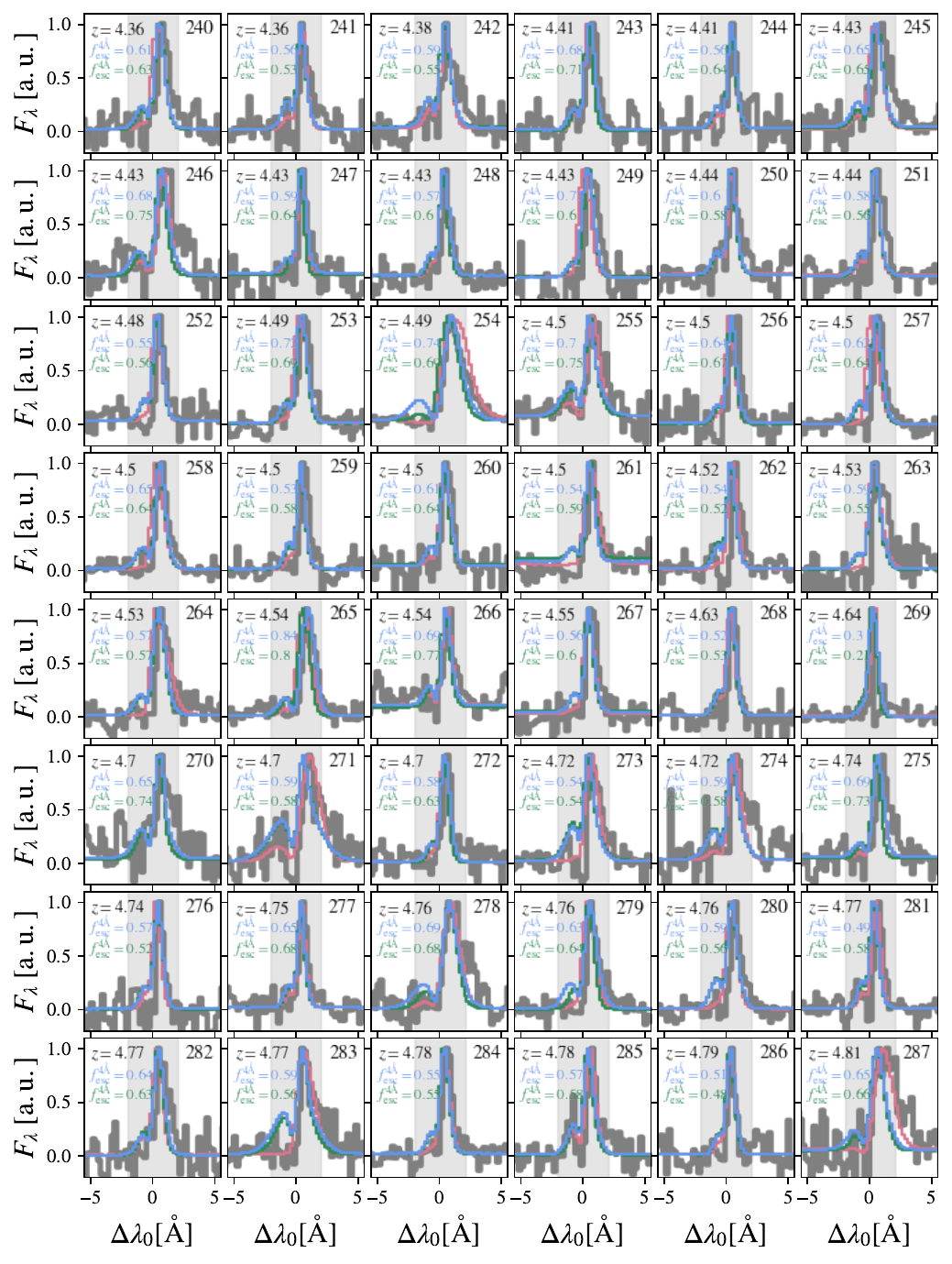}%
        \caption{ Same as Fig.~\ref{fig:LASD_0}, but for other observed \lya line profiles between $z=4.3$ and $z=4.8$. }
        \label{fig:LASD_5}
\end{figure*}
\begin{figure*} 
        \includegraphics[width=7.1in]{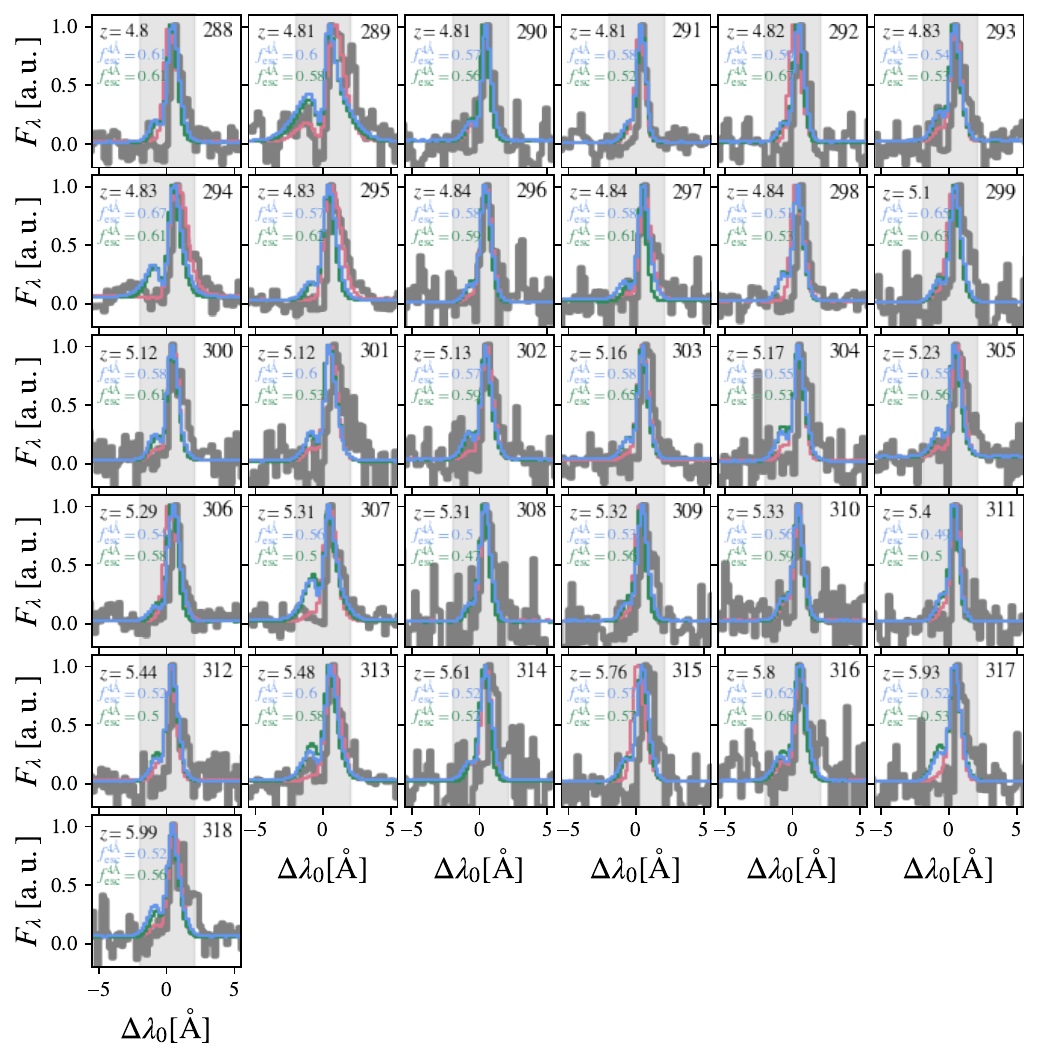}%
        \caption{ Same as Fig.~\ref{fig:LASD_0}, but for other observed \lya line profiles between $z=4.8$ and $z=6.0$. }
        \label{fig:LASD_6}
\end{figure*}

\begin{figure*} 
        \includegraphics[width=7.1in]{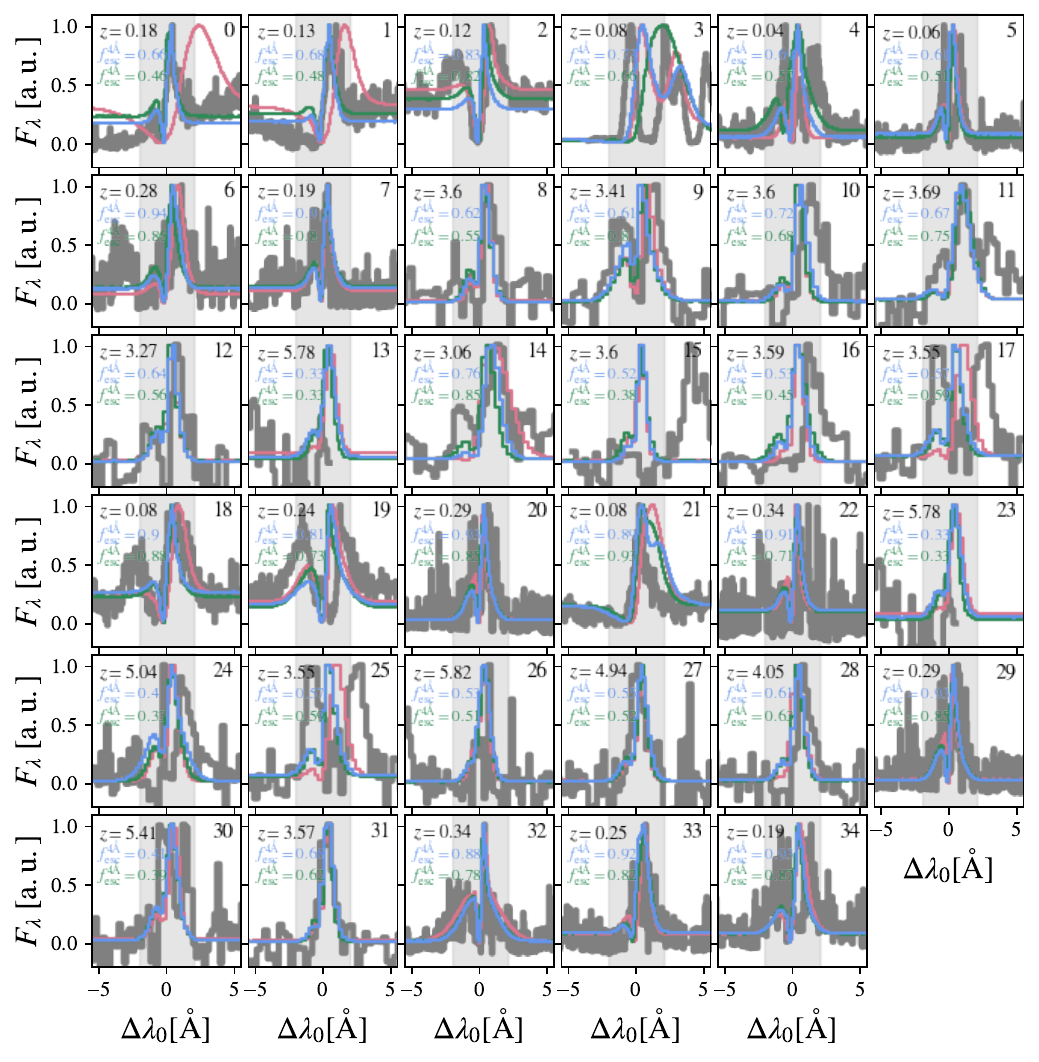}%
        \caption{ Same as Fig.~\ref{fig:LASD_0}, but for other observed \lya line profiles  that exhibit low \sn (e.g., 0 , 1 ), a steep continuum not modeled in \zelda (e.g., 0 , 1 ), or with more than two components (e.g., 3). }
        \label{fig:LASD_bad}
\end{figure*}

\end{document}